\documentclass[lettersize,journal]{IEEEtran}

\IEEEoverridecommandlockouts

\usepackage{amsfonts}
\usepackage[dvips]{graphicx}
\usepackage{times}
\usepackage{cite}
\usepackage{amsmath}
\usepackage{cases}
\usepackage{array}
\usepackage{dsfont}
\usepackage{amssymb}

\usepackage{makecell}
\usepackage{stfloats}
\usepackage{booktabs}
\usepackage{graphicx}
\usepackage{subfigure}
\usepackage{pdfpages}
\usepackage{footnote}
\usepackage{soul}
\usepackage{array}
\usepackage{multirow}
\usepackage{bm}
\usepackage{empheq}
\usepackage{amsthm}
\usepackage{algorithm}
\usepackage{algorithmicx}
\usepackage{algpseudocode}
\usepackage{color}
\usepackage{diagbox}
\usepackage{caption}
\theoremstyle{plain}

\theoremstyle{definition}

\usepackage[colorlinks,bookmarksopen,bookmarksnumbered,citecolor=blue, linkcolor=blue, urlcolor=blue]{hyperref}

\let\oldReturn\Return
\renewcommand{\Return}{\State\oldReturn}

\usepackage{textcomp}
\usepackage{xcolor}
\ifCLASSINFOpdf
\else
\fi

\begin{document}
	\title{Image Generation with Supervised Selection Based on Multimodal Features for Semantic Communications}
	\vspace{-4pt}
	
	\vspace{-25pt}\author{\IEEEauthorblockN{$\text{Chengyang Liang}$, $\text{Dong Li}, ~\IEEEmembership{Senior Member,~IEEE}$}
		\vspace{-22pt}
		\thanks{Chengyang Liang and Dong Li are with the School of Computer Science and Engineering, Macau University of Science and Technology, Macau, China (e-mail: 3240006992@student.must.edu.mo; dli@must.edu.mo).}}    
	\vspace{-20pt}

\maketitle
\begin{abstract}
	Semantic communication (SemCom) has emerged as a promising technique for the next-generation communication systems, in which the generation at the receiver side is allowed with semantic features' recovery. However, the majority of existing research predominantly utilizes a singular type of semantic information, such as text, images, or speech, to supervise and choose the generated source signals, which may not sufficiently encapsulate the comprehensive and accurate semantic information, and thus creating a performance bottleneck. In order to bridge this gap, in this paper, we propose and investigate a SemCom framework using multimodal information to supervise the generated image. To be specific, in this framework, we first extract semantic features at both the image and text levels utilizing the Convolutional Neural Network (CNN) architecture and the Contrastive Language-Image Pre-Training (CLIP) model before transmission. Then, we employ a generative diffusion model at the receiver to generate multiple images. In order to ensure the accurate extraction and facilitate high-fidelity image reconstruction, we select the "best" image with the minimum reconstruction errors by taking both the aided image and text semantic features into account. We further extend multimodal semantic communication (MMSemCom) system to the multiuser scenario for orthogonal transmission. Experimental results demonstrate that the proposed framework can not only achieve the enhanced fidelity and robustness in image transmission compared with existing communication systems but also sustain a high performance in the low signal-to-noise ratio (SNR) conditions.

\end{abstract}


\IEEEpeerreviewmaketitle

\vspace{-4pt}
\section{Introduction} 
\label{sec:Introduction}

\IEEEPARstart{T}{he} exponential increase in wireless data traffic, coupled with the growing complexity of communication tasks, have necessitated the development of more efficient and intelligent communication systems. Traditional Shannon-based communication frameworks \cite{Shannon1948}, while laying the foundation, encounter significant challenges in addressing the requirements of emerging applications such as augmented reality, autonomous vehicles, and the Internet of Things (IoT). These applications not only necessitate high data rates but also demand low latency, high reliability, and context-aware communication \cite{Gu,LiteSC}. Semantic communication (SemCom) fundamentally reorients the focus from the precise reproduction of signals to the transmission of the essential meaning or intent of the information \cite{Overview}. This paradigm shift aligns with the increasing recognition that, in many contemporary applications, the objective extends beyond mere signal reproduction to facilitating specific tasks or decisions based on the conveyed information. By emphasizing semantics, these systems have the potential to reduce redundancy, enhance the resource utilization, and improve the quality of service for end-users. In this dynamic landscape, SemCom has emerged as a promising paradigm for next-generation wireless communications, offering potential advancements in efficiency, reliability, and intelligence \cite{Rethink}.

Recent advancements in artificial intelligence, particularly in deep learning (DL) and natural language processing (NLP), have catalyzed significant progress in SemCom \cite{SCcarrier}. These technologies facilitate the extraction, representation, and reconstruction of semantic information with unprecedented effectiveness. Deep neural networks, characterized by their capacity to learn complex patterns and representations, have become essential in the development of semantic encoders that compress information while preserving its meaning \cite{TowardSC}. For example, Xie et al. proposed a deep learning-based SemCom system for text transmission that demonstrated greater robustness to channel variations and achieved a superior performance compared to traditional methods, while maintaining the message comprehension \cite{TextDeepSC}. Similarly, Wang et al. developed a proximal policy optimization-based reinforcement learning algorithm integrated with an attention mechanism. Furthermore, the trade-off between computational complexity and communication efficiency has been explored in recent works such as \cite{tradeoffsc, MECsc}, which discuss offloading and power-computation balancing in semantic-driven mobile edge computing and DL inference systems.

Image SemCom primarily aims to reconstruct image information that is both complete and of high quality while utilizing the minimal bandwidth for transmission. This approach ultimately results in a high peak signal-to-noise ratio (PSNR) and multi-scale structural similarity index measure (MS-SSIM) for the received image. Many of these frameworks utilize convolutional neural networks (CNN) to execute specific tasks. For instance, Zhang et al. proposed a predictive and adaptive deep coding SemCom framework for wireless image transmission, which transmits images with a minimal compression rate, thereby significantly enhancing the bandwidth efficiency \cite{PADC}.

On the other hand, generative artificial intelligence (GAI) has garnered significant attention for advancing SemCom systems due to its ability to generate and interpret contextually rich content intelligently. Recent studies highlight GAI's integration into SemCom to address challenges like noise mitigation, data compression, and transmission efficiency \cite{GANSC,CCDM,DiffusionSC,MMSC,NLJSCNSC}. Recent studies have also begun addressing semantic communication under considerations of privacy preservation, information bottleneck theory, and trustworthiness of GAI-driven systems. For instance, \cite{PrivacySC} proposes a privacy-aware semantic framework for task-oriented 6G, \cite{BottleneckSC} introduces an information bottleneck method for graph data, and \cite{trustworthinsessSC} explores explainability and control in GAI-based semantic communication. For instance, Generative Adversarial Networks (GANs) have been employed to counter signal distortion without relying on the channel state information (CSI), enhancing system resilience and efficiency \cite{GANSC}. Similarly, diffusion model (DM) has proven effective in refining and reconstructing transmitted data by suppressing the channel noise \cite{CCDM,DiffusionSC}. GAI also optimizes the semantic coding of images, enabling high-quality reconstruction at low bit rates while preserving semantic integrity and enhancing resilience against channel instability \cite{GAISurvey}. Furthermore, its multimodal data fusion capabilities exploit inter-modal correlations to reduce redundancy and improve transmission efficiency \cite{MMSC}. Beyond efficient data management, GAI empowers SemCom to execute complex tasks, crucial for applications like autonomous driving and smart cities, where timely and accurate decoding and reconstruction of vast data volumes are imperative for reliable decision-making \cite{GAIATA}.

Note that the majority of previous studies primarily concentrate on employing GAI to suppress the effects of the channel noise, or avoiding signal distortion to achieve a high accuracy, or using Generative adversarial networks (GANs) or DMs for image generation at the receiver, and thus lacking precise control at the semantic level. Thus, a natural question arises on how to fully leverage the semantic information of signals to enhance the perception and supervision of GAI-generated outcomes. This problem has garnered limited attention, with the exception of our previous work. In \cite{first}, we developed an end-to-end SemCom system for image transmission, which incorporated an image feature encoder in conjunction with a pre-trained classification extractor at the transmitter. At the receiver, we implemented a DM that generates images based on the transmitted classification information, subsequently refining the target image using the transmitted image features. Our findings demonstrate that this approach yields favorable results. However, our previous work only supervised the selection of a single image feature for the target image, potentially resulting in insufficient supervision for the potential semantic information. In scenarios where a substantial number of highly similar images are present in the generated results, our method may struggle to identify the most matching image, leading to suboptimal outcomes. In this paper, in order to bridge this gap, we propose a multimodal semantics (MultiSem) framework that integrates the semantics of both the signal image and the text to be transmitted. Building upon this framework, we design a semantic awareness multimodal semantic communication (MMSemCom) system. The contributions of this paper are as follows.

\begin{itemize}
	\item We propose a novel architecture for image-targeted MMSemCom. Distinct from other image SemCom systems, MMSemCom incorporates not only the semantic information derived from the target image but also the associated textual information to enhance the signal reconstruction. This integration of multimodal semantic information facilitates a more comprehensive extraction and understanding of the target, resulting in the improved fidelity of the transmitted image. Besides, we develop two methods for multimodal signal expression and transmission within this framework. The first method involves the transmission of both text and image after extracting their respective semantic modalities at the transmitter. At the receiver, the image is hierarchically recovered using the two types of semantics. The second method entails the extraction of both modal semantics at the transmitter, followed by their fusion into a unified multimodal semantic representation. At the receiver, the signal is restored using this fused multimodal semantics. Then we extend our framework to multiuser scenarios by using Walsh metrics for orthogonal transmission.
	\item We propose a new semantic encoder for MMSemCom within the transmitter, along with a novel indicator termed MultiSem in the encoder. The encoder is comprised of two components: the first is the feature extraction module, which utilizes CNN to extract features from the target image and employs Contrastive Language-Image Pre-Training (CLIP) to derive prompts associated with the image. The second component is the multimodal semantic encoding module, which incorporates two encoding methods for MultiSem. The first method involves the straightforward constituent of the semantics derived from image features and text features. The second method integrates the image features with the prompts extracted by CLIP, generating a new MultiSem representation through the application of an attention network.
	\item We propose a new semantic decoder that utilizes a DM within the receiver. The decoder employs the DM to generate the corresponding set of images, and adopts a selection module that incorporates two distinct selection methods, which effectively guarantee the error-free and efficient transmission of information. The first method adopts the sequential selection, while the second method considers simultaneous selection utilizing fused MultiSem. This pick-and-match mechanism ensures the maximization of semantic information and enhances the robustness of the transmission results.
	\item To ensure the effectiveness and universality of our proposed MMSemCom system, we conduct simulations utilizing a diverse array of datasets and varying signal-to-noise ratio (SNR) levels within an end-to-end single-user SemCom scenario. Under different simulation conditions, the system successfully achieves high-quality reconstruction of target images. 
\end{itemize}

The rest of this paper is organized as follows. In Sections \ref{sec:system model}, we show the general framework of our system. In Sections \ref{sec:architecture}, we show the transmission as well as reception process of image signals by a single-user end-to-end MMSemCom. In Section \ref{sec:Multi-User Scenario}, we show how to extend our proposed MMSemCom to a multi-user scenario. In Section \ref{sec:Complexity}, we derive and illustrate the communication overhead and computational complexity of the system. In Section \ref{sec:Experiments}, we present extensive simulation results to demonstrate the effectiveness and performance of our proposed system. Finally, Section \ref{sec:conclusion} concludes the results.

\vspace{-6pt}
\section{System Model}
\label{sec:system model}

\begin{figure*}[t]
\vspace{-8pt}
\centering
\includegraphics[scale=0.45]{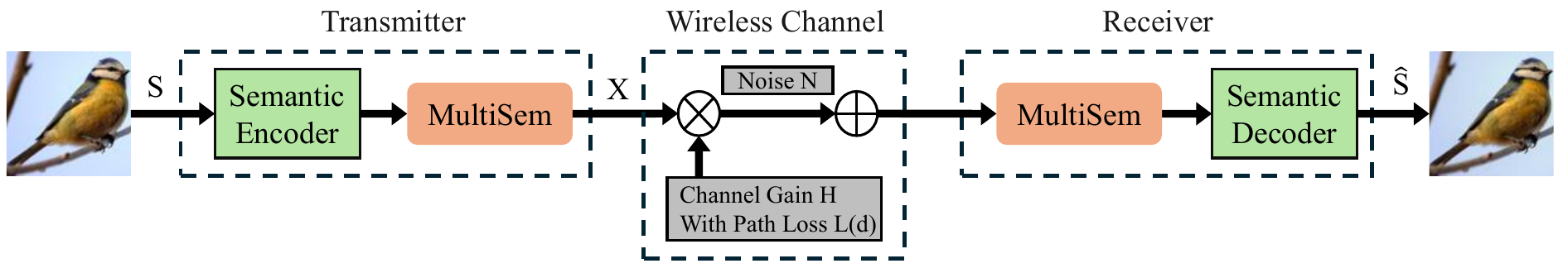}
\caption{Illustration of the proposed semantic communication system.}
\vspace{-9pt}
\label{fig:system}
\vspace{-2mm}
\end{figure*}

The system model under consideration comprises a transmitter, a receiver, and a wireless channel, wherein both the transmitter and the receiver share a common semantic knowledge base. As illustrated in Fig. \ref{fig:system}, the transmitter is tasked with the extraction and construction of multimodal semantic features, while the receiver is responsible for decoding and reconstructing the received semantic features. This entire process can be categorized into three primary stages: Semantic Encoding, Transmission, and Semantic Decoding.

The primary function of the transmitter is to extract multimodal semantic features from the target image signal, encompassing both image and text features. We denote the input signal as an image $S\in\mathbb{R}^{h\times w\times3}$, where $h$ represents the pixel height, $w$ indicates the pixel width, and 3 corresponds to the number of RGB color channels. Initially, the transmitter employs CNN and CLIP models to extract the semantic features of the image and text. Subsequently, these features are transferred to the multimodal semantic fusion module.

In our proposed semantic communication system, the semantic encoder at the transmitter is responsible for extracting the input image data into multi-modal semantic feature signals and mapping them into a dimensional representation space as $X = E_{\Theta}^{d_M}(S),$ where $S$ is the input raw image data, $X$ is the extracted multi-modal semantic representation. $E_{\Theta}(\cdot)$ represents the semantic encoder with a parameter $\Theta$, and $d_{M}$ denotes the dimension of the multimodal feature which can be summarized into three parts: image feature extraction module, text feature extraction module and fusion module.

In the image feature extraction module, we utilize a CNN architecture to perform this task. The multilayered structure of the CNN enables the progressive extraction of features, transitioning from low-level to high-level representations. In addition, the integration of pooling layers and non-linear activation functions within the CNN further enhances the abstraction and representation of features. As a result, the CNN architecture is chosen for its efficacy in extracting image features and its capacity to accurately represent the semantic content of images \cite{AlexNet}.
The process of extracting image features from a target image utilizing CNN can be articulated as $
 \mathcal{F}_{Image} = \Psi_{\eta_I}^{d_I}(S), $
where $\mathcal{F}_{Image} \in \mathbb{R}^{d_I}$ represents the features extracted from the image, $d_I$ denotes the dimensionality of these image features and $\Psi_{\eta_I}(\cdot)$ represents the image feature extraction function based on the CNN architecture.

In the text feature extraction module, we employ the CLIP model to generate corresponding prompts as text features from the input images. The CLIP model, which has been pre-trained on a large-scale dataset of image-text pairs, demonstrates an efficient capacity to comprehend image content and produce relevant textual descriptions \cite{CLIP}. The multimodal learning capabilities of CLIP enable it to establish semantic connections between images and text, thereby facilitating the generation of accurate text features. This methodology not only captures the visual information inherent in an image but also transforms it into a semantically rich textual representation, which is advantageous for the subsequent integration of multimodal semantics \cite{CLIPI2T}.

The procedure for extracting textual features from a target image utilizing the CLIP model can be articulated as $\mathcal{F}_{Text} = \Psi_{\eta_T}^{d_T}(S), $
where $\mathcal{F}_{Text} \in \mathbb{R}^{d_T}$ represents the prompts extracted from the image, $d_T$ denotes the dimensionality of the text features, $\Psi_{\eta_T}(\cdot)$ represents the text feature extraction function based on the CLIP.

Following the extraction of image and text features from the target image, we developed a multimodal semantic fusion module. This module facilitates the integration of image features and text features into a cohesive set of multimodal semantic features. This approach effectively synthesizes information from diverse modalities, thereby enhancing the expressive capacity of the features and producing more comprehensive and nuanced multimodal semantic representations.
The fusion of image features, denoted as $\mathcal{F}_{Image}$, and text features, represented as $\mathcal{F}_{Text}$, into multimodal semantic features, referred to as $\mathcal{F}_{MultiSem}$, can be articulated as follows

\begin{equation}
	\mathcal{F}_{MultiSem}=\Psi_{\eta_M}^{d_M}(\mathcal{F}_{Image},\mathcal{F}_{Text})=X,
\end{equation}
\noindent where $\mathcal{F}_{MultiSem} \in \mathbb{R}^{d_M}$ represents the fused multimodal semantic feature and $\Psi_{\eta_M}(\cdot)$ represents the fusion module based on the concatenation and attention mechanism.

During transmission, the prompts and multimodal semantics are conveyed through a wireless channel. Consequently, the received multimodal semantic signals  can be expressed as $\hat{y}=L(d)HX+N,$
\noindent where $H$ denotes the channel gain, $L(d)$ represents the path loss, where $d$ is the transmission distance, and $N$ is the noise at the receiver.

The primary function of the receiver is to reconstruct the image signal from the received MultiSem $\hat{y}$. The receiver employs a generative mechanism based on the DM to generate and select the image that most closely aligns with the transmitted semantic features. The DM is capable of progressively generating intricate image structures from simple noise distributions through a systematic denoising process. This methodology not only yields high-quality and diverse images but also effectively leverages conditional information to steer the generation process. In this study, we utilize conditional information to influence the type of images generated \cite{CDF}. When compared to other generative models, the DM demonstrates superior performance in terms of image quality and diversity, and its training process exhibits greater stability.
Initially, the receiver generates the image set $\hat{I}_{set}$ utilizing the conditional generation mechanism inherent in the DM as $\hat{I}_{set} = \xi_{\eta_D}^{h \times w}(\kappa),$
where $\xi_{\eta_D}(\cdot)$ generates the image set with the same length and width as the target image which is are respectively given by $h$ and $w$ based on the trained customized diffusion model, and $\kappa$ represents the shared dataset between the transmitter and the receiver.

Subsequently, the receiver systematically extracts the image features from each image within the generated set. These features are then integrated with the received text features, followed by a comparison with the received multimodal semantics. Ultimately, the receiver identifies the image $\hat{S}$ that exhibits the highest degree of correspondence, which is summarized as $\hat{S}=\xi_{\eta_S}(\hat{I}_{set};\hat{y}),$ where $\xi_{\eta_S}(\cdot)$ is the selection mechanism function for the received semantic signal $\hat{y}$ and the post-processing of the generated dataset $\hat{I}_{set}$ by the receiver.

\vspace{-2pt}
\section{architecture of encoder and decoder}
\label{sec:architecture}

\subsection{Semantic Encoder}
\label{sec:Semantic Encoder}
In this section, we will show the details of the transceiver design for the proposed SemCom system. In MMSemCom, effective extraction of image features is crucial for the system performance. In order to extract high-quality image features, this system employs a CNN model based on ResNet-50 in the semantic encoder, which belongs to the family of deep Residual Networks (ResNet), which overcome the gradient vanishing problem of deep networks through Residual Learning. ResNet-50 consists of 50 layers and utilizes skip connections, which allow gradients to propagate directly through the network, maintaining good training results even when the network is very deep. ResNet-50 is a classic architecture in CNN, which can efficiently extract high-level features from images for a variety of computer vision tasks \cite{ResNet}.

\subsubsection{Image Feature Extraction}
\label{sec:Image Feature}

In this system, the image to be transmitted is $S\in\mathbb{R}^{h\times w\times3}$. The image feature extraction module is based on the pre-trained ResNet-50 model using the pre-trained weights on ImageNet. This strategy can effectively utilize the powerful feature extraction capability of the pre-trained model and reduce the training time. We first load the pre-trained weights of the ResNet-50 model. This model has been trained on the ImageNet dataset and has a strong visual feature extraction capability.

To adapt the ResNet-50 model to the specific needs of SemCom, the model's final fully connected layer, originally used for classification, is removed. This modification enables the model to output high-level feature representations, capturing the essential semantic information from the image, instead of providing classification results. Denote the extracted features at the last convolutional block as $\mathbf{F}_{\mathrm{conv}}=f_{\mathrm{ResNet50}}(\mathbf{S})[:d_{\mathrm{conv}}]$, where 
$d_{\mathrm{conv}}$ represents the feature dimensionality of the original ResNet-50 output before any modification.

\begin{figure*}[htp]
	\vspace{-2mm}
	\centering
	\includegraphics[scale=0.4]{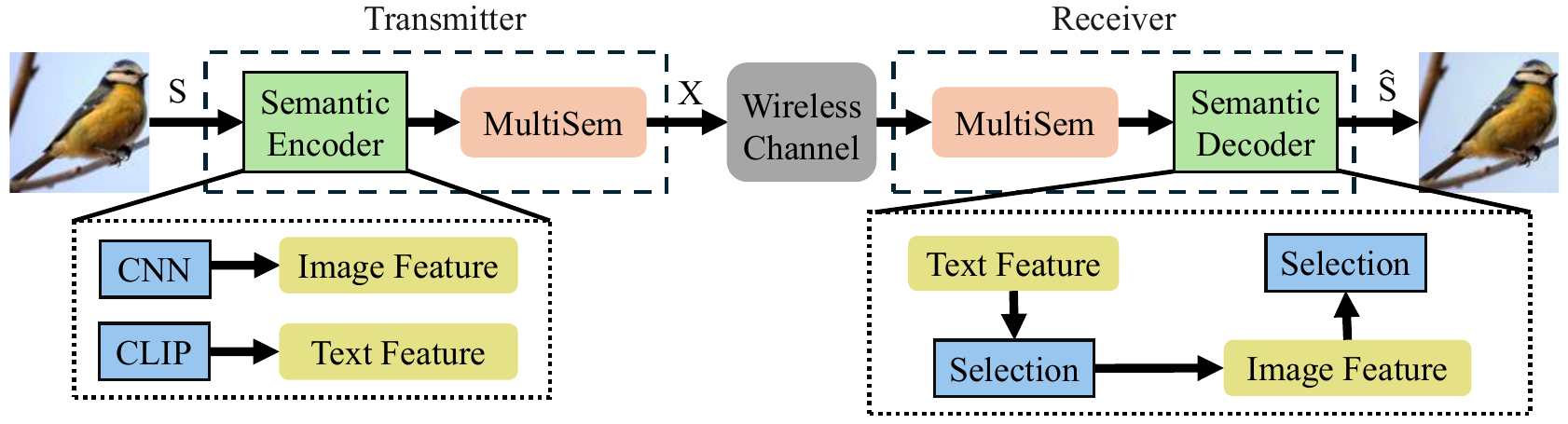}
	\caption{Illustration of the hierarchical sequential transmission method.}
	\label{fig:system_h1}
	\vspace{-1mm}
\end{figure*}

\begin{figure*}[htp]
	\vspace{-1mm}
	\centering
	\includegraphics[scale=0.4]{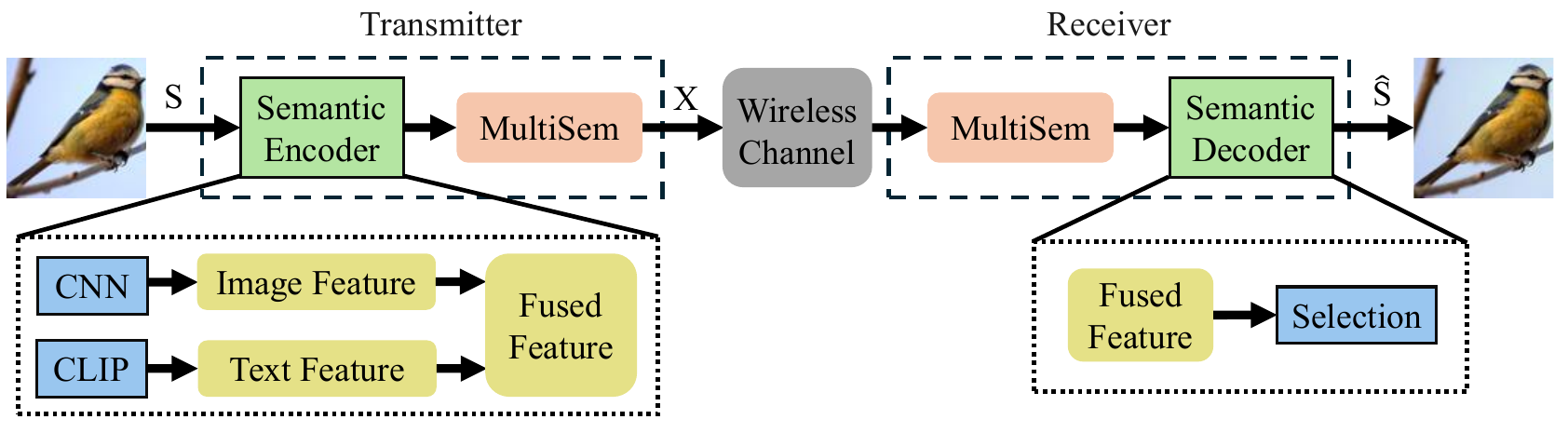}
	\caption{Illustration of the simultaneous transmission of fused semantic information.}
	\vspace{-9pt}
	\label{fig:system_s1}
	\vspace{-1mm}
\end{figure*}

To further reduce the dimensionality of $\mathbf{F}_{\mathrm{conv}}$ and facilitate multimodal feature fusion for transmission, we introduce two additional fully connected layers. The first fully connected layer transforms $\mathbf{F}_{\mathrm{conv}}$ into an intermediate feature vector of dimension $d_1$ with a non-linear ReLU activation to enhance the abstraction of essential features $\mathbf{F}_{d_1}=\sigma\left(\mathbf{W}_1\cdot\mathbf{F}_{\mathrm{conv}}+\mathbf{b}_1\right)$ , where $\mathbf{W}_1\in\mathbb{R}^{d_1\times d_{\mathrm{conv}}}, \mathbf{b}_1\in\mathbb{R}^{d_1}$, $d_{\mathrm{conv}}$, $d_1$ are the dimensionalities of the feature vectors after each transformation stage and 
$\sigma(\cdot)$ represents the ReLU activation function, enhancing feature non-linearity to support the efficient semantic expression required in communication tasks.

The intermediate feature vector $\mathbf{F}_{d_1}\in\mathbb{R}^{d_1}$ is then projected to a lower-dimensional semantic feature space, yielding the final feature vector $\mathcal{F}_{Image}\in\mathbb{R}^{d_{\mathrm{image}}}$ , which will be transmitted. This dimensionality reduction is achieved through another fully connected layer, allowing the preservation of the core semantic information while minimizing bandwidth consumption in the communication channel $\mathcal{F}_{Image}=\mathbf{W}_2\cdot\mathbf{F}_{d_1}+\mathbf{b}_2$, where
$\mathbf{W}_2\in\mathbb{R}^{d_{\mathrm{image}}\times d_1}, \mathbf{b}_2\in\mathbb{R}^{d_{\mathrm{image}}}$, $d_\mathrm{image}$ represents the final transmitted semantic feature vector's dimensionality. To summarize, the final extracted semantic feature vector 
$\mathcal{F}_{Image}$ for the image $S$ is formulated as

\begin{equation}
\mathcal{F}_{Image}=\mathbf{W}_2\cdot\sigma\left(\mathbf{W}_1\cdot\mathbf{F}_\mathrm{conv}+\mathbf{b}_1\right)+\mathbf{b}_2\in\mathbb{R}^{d_\mathrm{image}},
\end{equation}
\noindent where $\mathbf{W}_1,\mathbf{W}_2$ and $\mathbf{b}_1,\mathbf{b}_2$ are the learned weights and biases of the added fully connected layers.

\subsubsection{Text Feature Extraction}
\label{sec:Text Feature}
In MMSemCom, in addition to image features, prompts can be used as an additional semantic information to enhance the performance of the system. In order to realize the association between image and prompts, this system uses the CLIP model at the encoder for the extraction of prompts. The CLIP model is proposed by OpenAI with the aim of using contrastive learning to jointly train images and text on large-scale datasets. The advantage of this model is that it can learn to understand images from natural language descriptions, rather than relying only on limited artificial labels. CLIP is a powerful multimodal model that can process both images and text and map them into the same embedding space so that the similarity between the two can be computed \cite{CLIPKD}.

In this system, CLIP model is used to extract the most relevant prompts from a given image. First we need to load the pre-trained CLIP model and we use Vision Transformer as an image encoder to transform the image into a feature vector. Meanwhile, CLIP's text encoder can convert prompts into feature vectors and chooses the Text Transformer model as the text encoder. We use the “ViT-B/32” version of CLIP, which adopts a smaller model architecture and can reduce the amount of computation while ensuring the accuracy.

\begin{equation}
\begin{aligned}
	\mathcal{F}_{Text}&=CLIP(S,T)\\
               &=(ViT(S),Text-Transformer(T)).
 \end{aligned}
\end{equation}
Furthermore, to match the image and prompts, we provide a set of candidate prompts. The prompts are usually category and descriptive words related to the image content. $T=\{t_1,t_2,\ldots\,t_i\}$, where $t_i$ is the $ith$ prompts. CLIP's text encoder converts each prompt $t_i$ into a corresponding text feature vector $\mathcal{F}_{t_i}$, where $\mathcal{F}_{t_i} \in \mathbb{R}^{d_{text}}$ is the feature vector of the $ith$ prompt and $d_{text}$ is the dimensionality of each text feature vector. These text features also lie in the same dimensional embedding space as the image features.

To determine which prompts are most relevant to the image, the system calculates the cosine similarity between the image $S$ and each text feature $\mathcal{F}_{t_i}$ $C(S,\mathcal{F}_{t_i})=\frac{S\cdot\mathcal{F}_{t_i}}{\|S\|\|\mathcal{F}_{t_i}\|}$. The similarity is then converted to a probability distribution to represent the probability that the image is related to the prompt 

\begin{equation}
	P(t_i|S)=\frac{\exp(C(S,\mathcal{F}_{t_i}))}{\sum_j\exp(C(S,\mathcal{F}_{t_j}))}.
\end{equation}
Ultimately, the system selects the prompt that is most relevant to the image based on the similarity

\begin{equation}
	\mathcal{F}_{Text}={arg\,max}_{i}P(t_i|S) \in \mathbb{R}^{{n_{text}} \times {d_{text}}},
\end{equation}
where the number of text features, denoted by $n_{text}$, is equal to the number of text prompts.

\subsubsection{Description of MultiSem}
\label{sec:MultiSem}
In MMSemCom, to fully leverage the semantic information derived from image features and prompts, these two elements can be effectively integrated into a framework referred to as MultiSem. We have developed two distinct representations of MultiSem, which are illustrated in Fig. \ref{fig:system_h1} and Fig. \ref{fig:system_s1}. The first representation of MultiSem employs text prompts to encapsulate the text-level semantics of the target, while the image features represent the image-level semantics. These two forms of semantics are concatenated to create a multimodal semantic representation. During the transmission process, the system is required to convey the text semantics, which are represented by the prompt words, alongside the image semantics, represented by the image feature matrix. The second representation utilizes an attention mechanism to amalgamate the text prompts and image features into a singular semantic feature. This mechanism enables the prompts to execute attentional operations on the image features, thereby capturing the significant correlation between the prompts and the image semantics. In this transmission process, the system must transmit the fused feature matrix that results from the integration of the prompt words and the image feature matrix. Through this mechanism, the prompts direct the model's attention towards specific image regions or attributes that are described in the text. This approach enhances the semantic expressiveness of the system and facilitates efficient semantic transfer, even in environments characterized by the AWGN.

This paper proposes an enhanced cross-attention fusion model for the integration of image features $\mathcal{F}_{Image}$ and text features $\mathcal{F}_{Text}$. The proposed model employs a multi-head attention mechanism to facilitate the efficient fusion of image and text features. In particular, the image and text features are mapped to the same dimensional space, and cross-attention is then implemented through the multi-head attention mechanism. This enables the text features to attend to different parts of the image features. Subsequently, the features undergo further processing and fusion through residual joining, layer normalization and a feedforward network. First, we project $\mathcal{F}_{Image}$ and $\mathcal{F}_{Text}$ into the same fusion space dimension $d_{fusion}$ to generate the projection features $\mathcal{F}_{\mathrm{Image}}^{\prime}$ and $\mathcal{F}_{\mathrm{Text}}^{\prime}$

\begin{equation}
\begin{aligned}
\mathcal{F}_{Image}^{\prime}=W_{image}\mathcal{F}_{Image}+b_{image}, \\
\mathcal{F}_{Text}^{\prime}=W_{text}\mathcal{F}_{Text}+b_{text},
\end{aligned}
\end{equation}

\noindent where $W_{image}\in\mathbb{R}^{{d_{fusion}}\times d_{image}}$ and $W_{text}\in\mathbb{R}^{d_{fusion}\times d_{fusion}}$ are linear projection matrices  and $b_{text}$ are the bias terms. The dimensions of projected features $\mathcal{F}_{Image}^{\prime}$ and $\mathcal{F}_{Text}^{\prime}$ are denoted respectively as
$\mathcal{F}_{Image}^{\prime}\in\mathbb{R}^{d_{fusion}},\quad\mathcal{F}_{Text}^{\prime}\in\mathbb{R}^{n_{text}\times d_{fusion}}$

In the context of fusion space, the multi-head cross-attention mechanism is utilized to enhance the interaction between textual and image features, resulting in the generation of the attention output, referred to as $\mathcal{A}_{output}$. In multi-head attention, each head is tasked with calculating the attention weights and subsequently producing an output:

\begin{equation}
\mathcal{A}_{output}=Attention(\mathcal{F}_{Text}^{\prime},\mathcal{F}_{Image}^{\prime},\mathcal{F}_{Image}^{\prime}),
\end{equation}

\noindent where $Attention(\cdot)$ represents the multi-head attention operation, $\mathcal{F}_{Image}^{\prime}$ is the key and value and $\mathcal{F}_{Text}^{\prime}$ is the query. $\mathcal{A}_{output}$ is the shape of the $n_{text} \times d_{fusion}$. To maintain the feature consistency, we add residual connection and layer normalization to the output of cross-attention, and calculate the new feature $\mathcal{A}_{residual}=\phi(\mathcal{A}_{output}+\mathcal{F}_{Text}^{\prime})$, where $\phi(\cdot)$ represents the layer normalization operation. The residual connected features, denoted as $\mathcal{A}_{{residual}}$, are processed through a two-layer feed-forward network to facilitate further information fusion $\mathcal{F}_{final}=\sigma \times W_4(W_3\mathcal{A}_{residual}+b_3)+b_4$, where $W_3 \in \mathbb{R}^{4d_{{fusion}} \times d_{{fusion}}}$ and $W_4 \in \mathbb{R}^{d_{{fusion}} \times 4d_{{fusion}}}$ are the weight matrices of the feedforward network, $b_3$ and $b_4$ are the bias terms. The final fusion features are subsequently derived through the application of residual connections and layer normalization once more $\mathcal{F}_{nor}=\phi(\mathcal{F}_{final}+\mathcal{A}_{residual})$.

The text-dimension average pooling operation is ultimately employed to consolidate the final fused feature representations, denoted as $\mathcal{F}_{nor}$, into a singular fused feature vector $\mathcal{F}_{Fused}$

\begin{equation}
\mathcal{F}_{Fused}=\frac{1}{n_{text}}\sum_{i=1}^{n_{text}}\mathcal{F}_{{nor}}[i],
\end{equation}

\noindent where $\mathcal{F}_{Fused} \in \mathbb{R}^{d_{fusion}}$ denotes the resultant fused feature vector, which is applicable for SemCom as well as various subsequent tasks.

Note that traditional end-to-end semantic communications systems employing standard CNN, CLIP, and DM are not capable of significantly improving the robustness against channel degradation. The cross-attention fusion model we employ is a customized feature fusion approach. Its primary objective is to generate a highly recognizable and compact fused feature for the subsequent selection task. Through a learning process, the proposed mechanism dynamically assigns weights to the text prompts from CLIP and the image features from CNN, thereby maximizing the distance between matched and non-matched images in the semantic space. As shown later in Fig. \ref{fig:attention_heatmap}, the proposed attention mechanism effectively localizes the text concepts to their corresponding regions in the image, demonstrating its ability to extract efficient and reliable fusion features for the subsequent precise selection task.

\vspace{-8pt}
\subsection{Semantic Decoder}
\label{sec:Semantic Decoder}

In a manner analogous to the decoder function utilized in other communication systems, the semantic decoder module within the MMSemCom aims to reconstruct the original signal to the greatest extent possible based on the received multimodal semantic information. This process is executed in two stages:
\begin{itemize}
	\item First, we utilize the DM to conditionally generate a collection of images corresponding to the designated category. This procedure encompasses both the training phase and the generation of image samples within that specific category.
	\item Second, we propose two selection mechanisms for the distinct modalities of multimodal semantic expression. Utilizing these different selection mechanisms, the MultiSem of each image are compared with the received MultiSem. The MultiSem that exhibits the closest correspondence to the target of the semantic decoder is identified, and the images that fulfill the specified criteria are filtered from the image set generated in the first step, thereby serving as the restored image signal.
\end{itemize} 

\subsubsection{The Diffusion model generates the image set}
\label{sec:Diffusion}

In SemCom systems, generative models are particularly effective for transmitting high-quality images with minimal information. By learning the data's probability distribution, these models eliminate the need to transmit complete original data, requiring only key semantic information for the receiver to reconstruct high-fidelity images. This aligns seamlessly with the demands of efficient communication, as the model reconstructs original images from partial semantic data, significantly reducing the amount of transmitted raw data.

\begin{figure}[tb]
	\vspace{-6pt}
	\centering
	\includegraphics[scale=0.4]{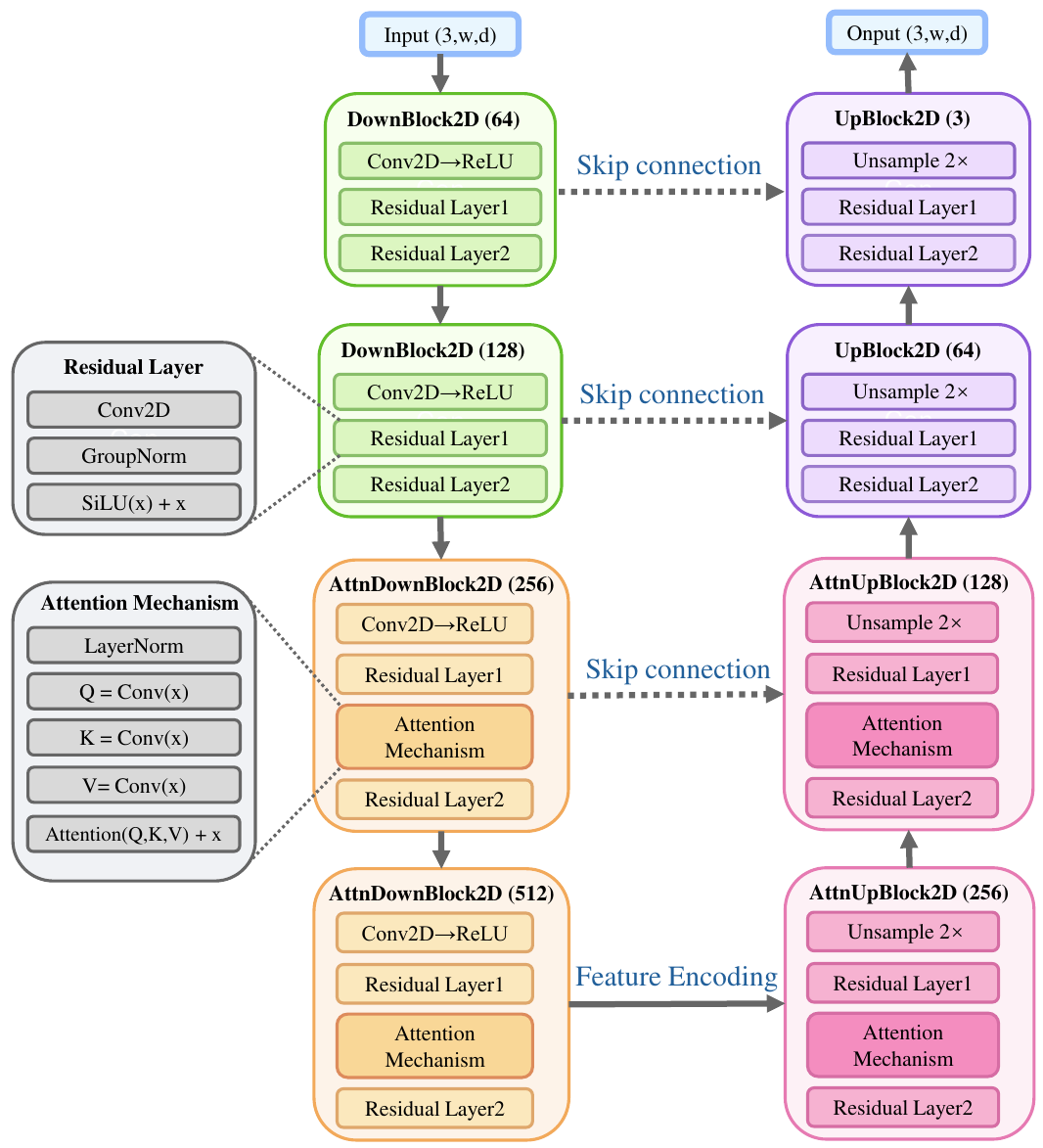}
	\caption{Demonstration of the diffusion network architecture designed for MMSemCom.}
	\vspace{-11pt}
	\label{fig:diffusion_model}
	\vspace{-2pt}
\end{figure}

In order to strengthen the feature capture ability of our DM for specific images, we propose a new network structure based on UNet. The architecture is specifically designed with enhanced hierarchical feature extraction and attention mechanisms to balance local detail preservation and global contextual modeling. As shown in Fig. \ref{fig:diffusion_model}, the model follows a symmetric Unet structure with four stages of downsampling and upsampling, where the downsampling progressively compresses spatial dimensions, and the upsampling reconstructs high-resolution features through learnable upsampling operations. Each down/up-block contains two residual layers, where skip connections mitigate gradient vanishing and strengthen feature reuse. Deeper blocks integrate self-attention mechanisms with model long-range dependencies. Attention modules are strategically placed in middle-to-deep layers. These modules first normalize features via LayerNorm, then compute query (Q), key (K), and value (V) through separate convolutional projections. The output is fused with the original features via residual connections, allowing the model to focus on semantically critical regions while maintaining spatial coherence. High-frequency details from the downsampling are directly propagated to the upsampling through cross-stage skip connections, compensating for the information loss during downsampling.

Specifically, generative models can produce high-quality images conditioned on semantic inputs (e.g., categories or partial features), with the generation process guided by conditional information (e.g., labels or descriptions) rather than relying solely on random noise. In MMSemCom, the condition $\mathbf{c}$ represents the image category, enabling the receiver to generate the corresponding image based on $\mathbf{c}$ using the generative model. This approach substantially reduces data transmission, enhancing the communication efficiency. The conditional generation model aims to learn the conditional distribution

\begin{equation}
	 p(\mathbf{x}|\mathbf{c})=\int p(\mathbf{x}|\mathbf{z},\mathbf{c})p(\mathbf{z}|\mathbf{c})d\mathbf{z}, 
\end{equation}

\noindent where $\mathbf{z}$ represents Gaussian noise, $\mathbf{c}$ is the condition information, and $\mathbf{x}$ is the generated image. In diffusion models, the generation process further refines this approach by learning to produce samples $\mathbf{x}$ from $\mathbf{z}$ and $\mathbf{c}$.

During DM training, the model incrementally adds the noise to a clean image and learns to recover the original image from the noisy versions. Starting with a clean image $\mathbf{x}_0$, noise is added iteratively to produce noisy images $\mathbf{x}_t$ at various time steps. The objective is to train the model to reconstruct $\mathbf{x}_0$ from these noisy images. The noise addition process is expressed as

\begin{equation}
 \mathbf{x}_t=\sqrt{\alpha_t}\mathbf{x}_0+\sqrt{1-\alpha_t}\mathbf{z},\quad\mathbf{z}\sim\mathcal{N}(0,\mathbf{I}), \end{equation}

\noindent where $t$ represents the time step, progressing from no noise to pure noise, $\mathbf{x}_t$ is the image at step $t$, $\alpha_t$ is a time-dependent weight decreasing with $t$, and $\mathbf{z}$ is Gaussian noise. During denoising, the model iteratively generates a less noisy image using the current noisy input and category labels, gradually reconstructing the target image.

The goal of training the DM is to enable the model to recover images from the noise and the process that can be defined by the following loss function. To effectively train the DM, we use a hybrid loss function that combines both L1 and L2 losses:

\begin{equation}
	\begin{aligned}
		\mathcal{L}(\theta) = &\mathbb{E}_{t,\mathbf{x}_0,\mathbf{z}} \bigg[ 
		\frac{1}{2} \|\mathbf{z}-\mathbf{f}_\theta(\sqrt{\alpha_t}\mathbf{x}_0 + 
		\sqrt{1-\alpha_t}\mathbf{z},t,\mathbf{c})\|_1 \\
		&+ \frac{1}{2} \|\mathbf{z}-\mathbf{f}_\theta(\sqrt{\alpha_t}\mathbf{x}_0 + 
		\sqrt{1-\alpha_t}\mathbf{z},t,\mathbf{c})\|_2^2 \bigg].
	\end{aligned}
\end{equation}

The goal of this loss function is to minimize the gap between the model's prediction noise and the true noise $\mathbf{z}$. By continuously optimizing this loss function, the model learns to progressively restore clean images from the noise, given the condition $\mathbf{c}$. The L1 loss ensures sharper edges and preserves fine details, while the L2 loss stabilizes training and enforces overall the structural consistency. By combining these two losses, we are able to balance the local detail preservation and the global coherence, leading to an improved image reconstruction performance.

\begin{algorithm}[ht]
	\caption{Conditional Diffusion Model Training Process}
	\label{alg:diffusion_training}
	\begin{algorithmic}[1]
		\Require Training dataset $\mathcal{D}$, Number of timesteps $T$, Noise schedule $\beta_t$
		\Ensure Trained diffusion model $\epsilon_\theta$
		
		\Function{Train}{$\mathcal{D}, T, \beta_t$}
		\State Initialize model parameters $\theta$
		\While{not converged}
		\State Sample $(x_0, \mathbf{c}) \sim \mathcal{D}$ 
		\State $t \sim \text{Uniform}(1,T)$ 
		\State $\epsilon \sim \mathcal{N}(0,I)$ 
		
		\State // Calculate noise scaling factors
		\State $\alpha_t = \prod_{s=1}^t (1-\beta_s)$
		\State $\sigma_t = \sqrt{1-\alpha_t}$
		
		\State // Add noise to input image
		\State $x_t = \sqrt{\alpha_t}x_0 + \sqrt{1-\alpha_t}\epsilon$
		
		\State // Predict noise and calculate loss
		\State $\hat{\epsilon} = \epsilon_\theta(x_t, t, \mathbf{c})$ 
		\State $\mathcal{L} = \|\epsilon - \hat{\epsilon}\|_2^2$
		
		\State Update $\theta$ using gradient descent on $\mathcal{L}$
		\EndWhile
		\Return $\epsilon_\theta$
		\EndFunction
	\end{algorithmic}
\end{algorithm}

In the generation process, the model begins with pure noise and iteratively refines it into a high-quality image through multiple denoising steps. In conditional diffusion models, category information is incorporated to ensure the generated image aligns with the specified category. This conditional generation process is expressed as

\begin{equation} 
	\mathbf{x}_{t-1}=\mathbf{f}_\theta(\mathbf{x}_t,t,\mathbf{c})+\sigma_t\mathbf{z}, 
\end{equation}

\noindent where $\mathbf{f}_\theta$ denotes the model's prediction of the denoised image, $\sigma_t$ is the noise standard deviation controlling the noise level, and $\mathbf{c}$ represents the category condition. This formulation leverages both the time-step $t$ and the conditional information $\mathbf{c}$ to guide the generative process effectively.

After $T$ iterations, when $t=0$, the image output by the model is the final image generated, denoted by $\hat{\mathbf{x}} = f_\theta(\mathbf{x}_1, 1, \mathbf{c})$ Here $\hat{\mathbf{x}}$ denotes the image generated by the model under condition $\mathbf{c}$ the image generated by the model under condition $\mathbf{c}$. When we sample the same condition $\mathbf{c}$ multiple times (starting from different initial noise $\mathbf{x}_T$), a series of images are generated. The set of generated images can be represented as

\begin{equation}
	\mathcal{X}_{\mathrm{gen}}=\{\hat{\mathbf{x}}^{(i)}\mid i=1,2,\ldots,N\},
\end{equation}

\noindent where $\hat{\mathbf{x}}^{(i)}$ is the $ith$ generated image and $N$ is the number of generated images. Each time an image is generated, a different initial noise $\mathbf{x}_T^{(i)}$ and the same condition $\mathbf{c}$ are used.

\subsubsection{Selection mechanism}
\label{sec:Selection mechanism}

In this section, we introduce a selection mechanism aimed at identifying the most semantically similar image from a generated image set, denoted as $\mathcal{X}_{\mathrm{gen}}$, through a comparison with the received MultiSem. We have developed two distinct selection mechanisms tailored for different multimodal semantic expressions, as illustrated in Fig. \ref{fig:system_h1} and Fig. \ref{fig:system_s1}. The first mechanism, termed sequential selection, involves an initial filtering of the image set based on text semantics, followed by a further refinement using image features to achieve a more precise identification of the target image. The second mechanism, referred to as simultaneous selection, derives the target image by filtering the image set through the application of fusion semantics, which integrates both text and image semantics.

First, we will elucidate the process of sequential selection as illustrated in Fig. \ref{fig:system_h1}. In this framework, the transmitted text prompts and image features are denoted as $\tilde{\mathcal{F}}_{Text}$ and $\tilde{\mathcal{F}}_{\text{Image}}$, respectively. Let us consider the set of generated images to be selected, denoted as $\mathcal{X}_{\text{gen}}$. For each image $\hat{\mathbf{x}} \in \mathcal{X}_{\text{gen}}$, a CNN is employed to extract the image features $\mathcal{F}_{\text{Image}}(\hat{x})$. Concurrently, the CLIP model is utilized to extract the corresponding text features $\mathcal{F}_{\text{Text}}(\hat{x})$. Subsequently, the average distance between the extracted text feature and the received text feature $\tilde{\mathcal{F}}_{\text{Text}}$ is computed to assess their semantic similarity. The distance function is defined as follows
\vspace{-1.5mm}
\begin{equation}
\begin{aligned}
{Dist}_{text}\Big(\mathcal{F}_{Text}(\hat{\mathbf{x}}), \tilde{\mathcal{F}}_{Text}\Big) &= \\
\frac{1}{p} \sum_{i=1}^{p} \Bigg( 
\gamma\Big(\mathcal{F}_{Text}(\hat{\mathbf{x}})_i, \tilde{\mathcal{F}}_{Text}\Big) &+ \xi\Big(\mathcal{F}_{Text}(\hat{\mathbf{x}})_i, \tilde{\mathcal{F}}_{Text}\Big) 
\Bigg).
\end{aligned}
\end{equation}

\noindent In this context, $\gamma(\cdot)$ and $\xi(\cdot)$ denote the cosine distance and Euclidean distance, respectively, while $p$ represents the total number of images within the selected image set. The ${Dist}_{text}$ value calculated for each image serves to identify the top $q$ images that exhibit the highest similarity in features to the provided text prompts. The computations for $\gamma(\cdot)$ and $\xi(\cdot)$ are conducted as follows

\begin{equation}
\gamma(\cdot)=\frac{A \cdot B}{\|A\|\|B\|}=\frac{\sum_{i=1}^nA_iB_i}{\sqrt{\sum_{i=1}^nA_i^2}\sqrt{\sum_{i=1}^nB_i^2}},
\end{equation}

\begin{equation}
\xi(\cdot)=\sqrt{\sum_{i=1}^N(A_{i}-B_{i})^2}.
\end{equation}

\noindent For the top $q$ candidate images, the distance between the image features of each candidate image and the received image features is calculated. Specifically, the received image feature is denoted as $ \tilde{\mathcal{F}}_{\text{Image}} $, while the feature of the candidate image is represented as $ \mathcal{F}_{\text{Image}}(\hat{x}) $. The image feature distance is defined as follows
\vspace{-1.5mm}
\begin{equation}
\begin{aligned}
{Dist}_{image}\Big(\mathcal{F}_{Image}(\hat{\mathbf{x}}), \tilde{\mathcal{F}}_{Image}\Big) &= \\
\frac{1}{q} \sum_{i=1}^{q} \Bigg( \gamma\Big(\mathcal{F}_{Image}(\hat{\mathbf{x}})_i, \tilde{\mathcal{F}}_{Image}\Big) &+ \xi\Big(\mathcal{F}_{Image}(\hat{\mathbf{x}})_i, \tilde{\mathcal{F}}_{Image}\Big) \Bigg).
\end{aligned}
\end{equation}
The integration of cosine distance and Euclidean distance is employed to assess the similarity of image features. The image exhibiting the closest features to the input image feature is subsequently identified as the final similar image result.

Then, we present the simultaneous selection mechanism illustrated in Fig. \ref{fig:system_s1}. In the initial step, we extract text prompts and image features for all images within the image set. For each image in the image set, denoted as $\hat{\mathbf{x}} \in \mathcal{X}_{\mathrm{gen}}$, the corresponding extracted text prompt is represented as $\mathcal{F}_{Text}(\hat{\mathbf{x}})$, while the image feature is denoted as $\mathcal{F}_{Image}(\hat{\mathbf{x}})$. The received MultiSem is represented as $\tilde{\mathcal{F}}_{Fused}$. We employ the cross-modal attention fusion function $Attention(\cdot)$ to coalesce the image feature $\mathcal{F}_{Image}(\hat{\mathbf{x}})$ and the text prompt $\mathcal{F}_{Text}(\hat{\mathbf{x}})$ into the MultiSem $\mathcal{F}_{Fused}(\hat{\mathbf{x}})$

\begin{equation}
\mathcal{F}_{Fused}(\hat{\mathbf{x}})=Attention(\mathcal{F}_{Text}(\hat{\mathbf{x}}),\mathcal{F}_{Image}(\hat{\mathbf{x}})).
\end{equation}
We further calculate the cosine distance and Euclidean distance between the multimodal feature $\mathcal{F}_{Fused}(\hat{\mathbf{x}})$ and the received multimodal feature $\tilde{\mathcal{F}}_{Fused}$ for each generated image $\hat{x}$. These distances are then aggregated to form a comprehensive similarity measure. The distance is defined as follows:

\begin{equation}
\begin{aligned}
{Dist}_{fused}\Big(\mathcal{F}_{Fused}(\hat{\mathbf{x}}), \tilde{\mathcal{F}}_{Fused}\Big) &= \\
\frac{1}{p} \sum_{i=1}^{p} \Bigg( 
\gamma\Big(\mathcal{F}_{Fused}(\hat{\mathbf{x}})_i, \tilde{\mathcal{F}}_{Fused}\Big) &+ \xi\Big(\mathcal{F}_{Fused}(\hat{\mathbf{x}})_i, \tilde{\mathcal{F}}_{Fused}\Big) 
\Bigg).
\end{aligned}
\end{equation}

\noindent In this context, $\gamma(\cdot)$ and $\xi(\cdot)$ denote the cosine and Euclidean distances, respectively. Let $p$ represent the total number of images within the image set. The image exhibiting the smallest distance is designated as the most semantically similar to the target reference image. This selection mechanism facilitates the efficient identification of the generated image that most closely aligns with the multimodal semantic features of the target image, thereby ensuring a meaningful and accurate retrieval process.


\vspace{-4pt}
\section{Multi-User Scenario}
\label{sec:Multi-User Scenario}

Extending our proposed MMSemCom framework to multi-user scenarios has significant practical implications and technical advantages. Modern wireless networks must support a large number of users transmitting media content concurrently, resulting in crowded spectrum resources. The core concept of MMSemCom is to transfer compact semantic features rather than large raw image data, which significantly reduces the bandwidth requirements for each user. This capability is essential for multi-user systems.

Moreover, one of the primary challenges in multiuser environments is inter-user interference, which manifests similarly to channel noise. Our MMSemCom architecture exhibits inherent robustness to noise and interference through its unique generation and selection decoding mechanism. Even when the transmitted semantic features are moderately distorted, the selection mechanism maintains a high probability of identifying the image that closely resembles the original semantics, thereby effectively mitigating the impact of inter-user interference on the final image quality. Consequently, the low bandwidth requirements and strong robustness of MMSemCom position it as a highly promising solution for addressing the challenges of multi-user communication.

In a multi-user scenario, we consider a scenario involving that there are $n$ users, each characterized by a multimodal semantic feature vector $\mathbf{v}_i \in \mathbb{R}^k$. To facilitate efficient resource sharing among multiple users and to maintain the orthogonality of user signals, we propose the utilization of the Walsh code matrix for the encoding and decoding of the users' feature vectors.

In order to facilitate multi-user communication, we initially construct a Walsh code matrix denoted as $\mathbf{W}_d$, which has dimensions of $d \times d$ and adheres to the property of row-column orthogonality.
 $\mathbf{W}_d\mathbf{W}_d^T=d\cdot\mathbf{I}_d$, where $\mathbf{I}_d$ denotes the identity matrix of dimension $d \times d$. This implies that the rows and columns of the matrix are orthogonal to one another, with each row $\mathbf{w}_i \in \mathbb{R}^d$ corresponds to representing a distinct user $i$. The dot product between different rows is zero, expressed mathematically as $\mathbf{w}_i \mathbf{w}_m^T = 0$ for $i \neq m$, while the dot product of a row with itself is equal to $d$, i.e., $\mathbf{w}_i \mathbf{w}_i^T = d$. This orthogonality property guarantees that user signals remain non-interfering during the processes of encoding and decoding in a multi-user environment.

At the transmitter, the feature vector $\mathbf{v}_i$ associated with user $i$ is segmented and encoded using the corresponding Walsh code. It is assumed that the feature vector $\mathbf{v}_i$ offer each user $i$ is partitioned into $b$ segments, each comprising $d$

\begin{equation}
	\mathbf{v}_i=\begin{bmatrix}\mathbf{v}_{i,1}, & \mathbf{v}_{i,2}, & \cdots, & \mathbf{v}_{i,b}\end{bmatrix}^\top,\quad\mathbf{v}_{i,j}\in\mathbb{R}^d, \label{con:22}
\end{equation}

\noindent For each block $\mathbf{v}_{i,j}$, the coded signal is generated by multiplying it with the corresponding Walsh code line  $\mathbf{s}_{i,j}=\mathbf{v}_{i,j}\cdot\mathbf{w}_i$. The encoded signal $\mathbf{s}_i$ for the entire user $i$ is denoted as follows

\begin{equation}
	\mathbf{s}_i=\begin{bmatrix}\mathbf{v}_{i,1}\cdot\mathbf{w}_i, & \mathbf{v}_{i,2}\cdot\mathbf{w}_i, & \cdots, & \mathbf{v}_{i,b}\cdot\mathbf{w}_i\end{bmatrix}^\top,\quad\mathbf{s}_i\in\mathbb{R}^b, \label{con:23}
\end{equation}

\noindent For $n$ users, the sum of the coded signals for all users is: $\mathbf{S}=\sum_{i=1}^n\mathbf{s}_i$.

Assuming that the communication channel is affected by AWGN, the total transmitted signal can be represented as: $\mathbf{y}=H * \mathbf{S}+N$, where $H$ denotes the channel gain, $N \sim \mathcal{N}(0, \sigma^2 \mathbf{I}_d)$ denotes the Gaussian noise, and $\sigma^2$ represents the noise power.

At the receiving end, the received signal $\mathbf{y}$ comprises the superimposed signals from all users. To isolate the signal of user $i$, the receiving end multiplies the total signal $\mathbf{y}$ by the Walsh code vector $\mathbf{w}_i$ that corresponds to user $i$

\begin{equation}
	\mathbf{w}_i^T\mathbf{y}=\mathbf{w}_i^T\left(H*\sum_{i=1}^n\mathbf{s}_i+\mathbf{n}\right). \label{con:24}
\end{equation}

The application of the orthogonality property of Walsh codes facilitates the elimination of signals from non-correlated users, resulting in

\begin{equation}
	\mathbf w_i^T\mathbf y=H*d\begin{bmatrix}\mathbf v_{i,1}\\\mathbf v_{i,2}\\\dots\\\mathbf v_{i,b}\end{bmatrix}+\mathbf w_i^T\mathbf n=H*d\mathbf v_i+\mathbf w_i^T\mathbf n. \label{con:25}
\end{equation}

The signal of user $i$ can be expressed as a result of the orthogonality of Walsh codes

\begin{equation}
	\hat{\mathbf{v}}_i\approx\frac{1}{H*d}\mathbf{w}_i^T\mathbf{y}. \label{con:26}
\end{equation}
\noindent The term $\mathbf{w}_i^T \mathbf{n}$ represents the noise component, and it is observed that the impact of noise diminishes as the variable $d$ increases.

\vspace{-8pt}
\section{Communication Overhead and Computational Complexity}
\vspace{-2pt}
\label{sec:Complexity}

In this section, we elucidate the bandwidth requirements necessary for the two different multimodal semantic transfer methods, as well as the transmission of the selection mechanism and the varying computational complexities associated with each mode.

\vspace{-2pt}
\subsection{Communication Overhead}
The first method of the multimodal semantic transceiver, which facilitates the transmission of image features and text prompts alongside sequential selection within the selection mechanism, is delineated herein. This process necessitates the transmission of text prompts and image features from the transmitter to the receiver. Initially, filtering is conducted based on the feature vector of the text prompts. Following the selection of text features, image feature selection is subsequently employed. The text features, denoted as features $\mathcal{F}_{Text} \in \mathbb{R}^{n_{text} \times d_{text}}$ are generated by a text encoding model, such as CLIP, and possess dimensions of $n_{text} \times d_{text}$. Assuming that the features are represented as 32-bit floating-point numbers, with each feature value occupying 4 bytes, the transfer cost : $C_\mathrm{Text}=n_{text} \times d_{text} \times 4 $. The image feature $\mathcal{F}_{Image} \in \mathbb{R}^{d_{image}}$ is produced by an image encoding model, such as ResNet-50, which operates in the dimensional space of ResNet-50 with dimension $\mathbb{R}^{d_{image}}$. The transfer overhead associated with this feature is given by $C_\mathrm{Image}=d_{image}\times4$. In conclusion, the total communication cost in bytes for the first selection method is as follows

\begin{equation}
C_1=C_\mathrm{Text}+C_\mathrm{Image}=4 \times (n_{text} \times d_{text} + d_{image}).
\end{equation}

In the second approach, the multimodal semantic transceiver is required to transmit only the fused multimodal semantics. The dimensionality of the multimodal semantics is denoted as $\mathbb{R}^{d_{fusion}}$. Correspondingly, the number of bytes necessary for communication is as $C_2=4 \times d_{fusion}.$

\vspace{-6pt}
\subsection{Computational Complexity}

In this analysis of complexity, we will systematically evaluate the complexity associated with image feature extraction, text feature extraction, the fusion process, the sampling of images using diffusion models, and two selection mechanisms. This approach aims to elucidate the complexity expressions of two multi-modal semantic transceivers.

The process of extracting image features utilizing the ResNet-50 architecture, followed by dimensionality reduction through two fully connected layers, can be mathematically represented as $\mathcal{O}_{image}=\mathcal{O}(h \cdot w \cdot C_{res} \cdot K^2 ) + \mathcal{O}(d_{conv} \cdot d_1) + \mathcal{O}(d_1 \cdot d_{image})$. In this context, $C_{res}$ denotes the number of channels in the convolutional layers of the model, while $K$ represents the size of the convolutional kernel. The process of text feature extraction employs the CLIP model to derive feature vectors from text prompts. The complexity of the CLIP model is primarily influenced by the characteristics of the pre-trained model $\mathcal{O}_{text}=\mathcal{O}(n_{text} \cdot d_{text}^2)$.
In the attention mechanism, we initially apply linear transformations to the image and text features independently. The complexity associated with this process is $\mathcal{O}(d_{image}\cdot d_{fusion}+n_{text} \cdot d_{text} \cdot d_{fusion})$. 

The complexity of the multi-head attention mechanism primarily arises from three key steps: linear transformation (which involves the generation of queries, keys, and values), the computation of attention scores, and the execution of a weighted summation of the results. The dimension of each attention head is defined as $d_{\mathrm{head}}=\frac{d_{fusion}}{n_{heads}}$, where $n_{heads}$ represents the number of attention heads. Consequently, the complexity of the multi-head attention mechanism can be expressed as $\mathcal{O}(d_{\mathrm{head}} \cdot n_{heads})=\mathcal{O}(d_{fusion})$. The complexities associated with residual connections and layer normalization are relatively minor and can be approximated as linear time complexity $\mathcal{O}(d_{fusion})$. The feed-forward neural network comprises two linear layers, each of which scales the dimension by a factor of four before subsequently reducing it. The complexity of this component is given by $\mathcal{O}(d_{fusion}\cdot(4\cdot d_{fusion}))=\mathcal{O}(d_{fusion}^2)$. By aggregating the complexities of the aforementioned components, the overall complexity of the fusion mechanism can be summarized as  $\mathcal{O}_{Attention}=\mathcal{O}(d_{image}\cdot d_{fusion}+n_{text} \cdot d_{text} \cdot d_{fusion})+\mathcal{O}(d_{fusion})+\mathcal{O}(d_{fusion}^2)$. 

In the phase of sampling using the DM to generate image sets, the complexity is primarily determined by the model architecture and the number of generation steps. Our model architecture predominantly employs the U-Net model \cite{Unet}. During the image sampling process to create the image set, $p$ images are generated, with each image having dimensions $h \times w$ and comprising three RGB channels. The overall complexity of the generation process can be expressed as $\mathcal{O}_{diffusion}=\mathcal{O}(p\cdot T\cdot h\cdot w\cdot 3 \cdot L \cdot K^2)$, where $T$ represents the number of inference steps in the diffusion process, and $L$ denotes the number of convolutional layers in the U-Net model.

Under the two selection mechanisms, the first sequential mechanism initially employs text prompts to filter $q$ images from an entire image set containing $p$ images. Subsequently, it utilizes image features for further filtering. The computational complexity is $\mathcal{O}_{seq}=\mathcal{O}\Big(p \cdot \mathcal{O}(n_{text} \cdot d_{text}) + \mathcal{O}(q\cdot d_{image})\Big)$. The complexity of the second simultaneous selection mechanism to filter $p$ images in the image set using only fused features is $\mathcal{O}_{sim}=\mathcal{O}(p\cdot d_{fusion})$.

The first multimodal SemCom method necessitates the extraction of textual prompts and image features at the source. At the receiving end, a diffusion process is employed to generate a set of $p$ images. Within this image set, the textual features of all images are extracted, followed by the filtering of $q$ images. Subsequently, the image features of these $q$ images are extracted and subjected to further filtering. The overall complexity of this process is as follows

\begin{equation}
(p+1) \cdot \mathcal{O}_{text} + (q+1) \cdot \mathcal{O}_{image} + \mathcal{O}_{diffusion} +\mathcal{O}_{seq}.
\end{equation}

In the second multimodal SemCom method, the text prompt words and image features are extracted and fused at the transmitter's end prior to transmission. At the receiving end, a similar extraction and fusion process is performed on the $p$ images generated by diffusion to identify the most consistent images. The overall complexity is as follows

\begin{equation}
\begin{aligned}
(p+1) &\cdot ( \mathcal{O}_{text} +  \mathcal{O}_{image} ) \\
&+ \mathcal{O}_{diffusion} + p \cdot \mathcal{O}_{Attention} +\mathcal{O}_{sim}.
\end{aligned}
\end{equation}

\section{Numerical Experiments}
\label{sec:Experiments}
In this section, we conduct simulation results on the proposed MMSemCom across a variety of data sets to prove the high efficiency and high fidelity performance of our proposed SemCom system.

\vspace{-8pt}
\subsection{Experiments Setting}

\subsubsection{Dataset}

We evaluate the proposed MMSemCom system using a variety of datasets that feature different image resolutions and semantic categories to thoroughly assess its performance and generalization capabilities. Our training datasets include the complete CIFAR-100 and STL-10 image sets, as well as a randomly selected subset of 100 categories from the ImageNet-256 dataset. CIFAR-100 consists of 60,000 color images with a resolution of $32 \times 32$ pixels, spanning 100 fine-grained object classes, each containing 600 images. STL-10 comprises 13,000 labeled images across 10 classes, with a higher resolution of $96 \times 96$ pixels, providing more visual detail than CIFAR-100. The ImageNet-256 subset we utilize includes $256 \times 256$ high-resolution images drawn from diverse object classes, offering a richer distribution and broader coverage of visual semantics for model training.

\subsubsection{Parameters Setting}

We utilize the aforementioned dataset to train the DM implemented in the receiver for a total of 50,000 epochs. The training process is accelerated through the use of two Nvidia A100 GPUs. In terms of training configuration, we set the batch size to 64 and employed a learning rate of $2 \times10^{-4}$ with the AdamW optimizer. The diffusion process is directed by a UniPC (Unified Predictor-Corrector) multistep scheduler, which serves as a high-order solver optimized for rapid sampling in diffusion models \cite{UniPC}. We configure the number of diffusion timesteps to 1,000, which strikes an effective balance between the quality of generation and computational efficiency.

\vspace{-8pt}
\subsection{Demonstration of an end-to-end MMSemCom}

In this section, we present two distinct approaches to multimodal semantic representation in multiple channel models, and we consider the Rayleigh channel model under path loss with AWGN. In the AWGN channel,  $N$ is the AWGN noise with $N \sim \mathcal{N}(0, \sigma_T^2)$ and $\sigma_T^2$ denoting the noise variance. In the Rayleigh channel with path loss, in Eq. 2, $H\sim\mathcal{CN}(0,\sigma_{h}^{2})$ corresponds to a Rayleigh fading channel. The path loss denotes as $L(d)=\sqrt{d^{-\alpha}}$, where $d$ represents the transmission distance, and $\alpha$ is the path loss exponent that characterizes the attenuation of the signal over distance. In out experiment, $\alpha$ is set to 2, corresponding to the free-space path loss model. The trained DM is utilized by both the transmitter and the receiver. We first illustrate the hierarchical sequential transmission mode, as depicted in Fig. \ref{fig:system_h1}. For the purposes of this discussion, we consider the transmitted image to be representative of the bird class from the STL-10 dataset, specifically the bird image displayed on the left side of Fig. \ref{fig:system_h1}.

We extract the image feature vector using ResNet-50, setting the dimension of $d_{image}$ to 32. The text prompts are obtained through the CLIP model, with the number of prompts $n_{text}$ equal to 3. Each prompt has a dimension $d_{fusion}$ of 512. Following the extraction of prompts from the image depicted in Fig. \ref{fig:system_h1}, the resulting prompts are `bird', `yellow', `on a tree'.

In the receiver, we train the DM for 50,000 epochs utilizing the same dataset, sampling $p=5000$ images at the inference time step $T=40$. Subsequently, we employ the text prompts to initially filter the nearest $q=100$ images from the 5000 sampled images, followed by a second filtering process using the features of the received image. The image that exhibits the smallest distance is identified as the reconstructed target signal.

The simultaneous transmission method illustrated in Fig. \ref{fig:system_s1} is presented. The conditions remain consistent with the previously stated assumptions. In the simultaneous transmission process, we extract the textual prompts and image features from the original image. Subsequently, these elements are fused into a fusion feature matrix with a dimensionality of $d_{fusion}=64$. This matrix is then transmitted to the receiver. At the receiver , an image set comprising 5000 images is generated using the DM. The received fusion feature matrix is employed for filtering purposes. The resulting image exhibits the minimum distance from the received fusion feature following the selection process.

\begin{figure}[tb]
	\vspace{-2pt}
	\centering
	\includegraphics[scale=0.185]{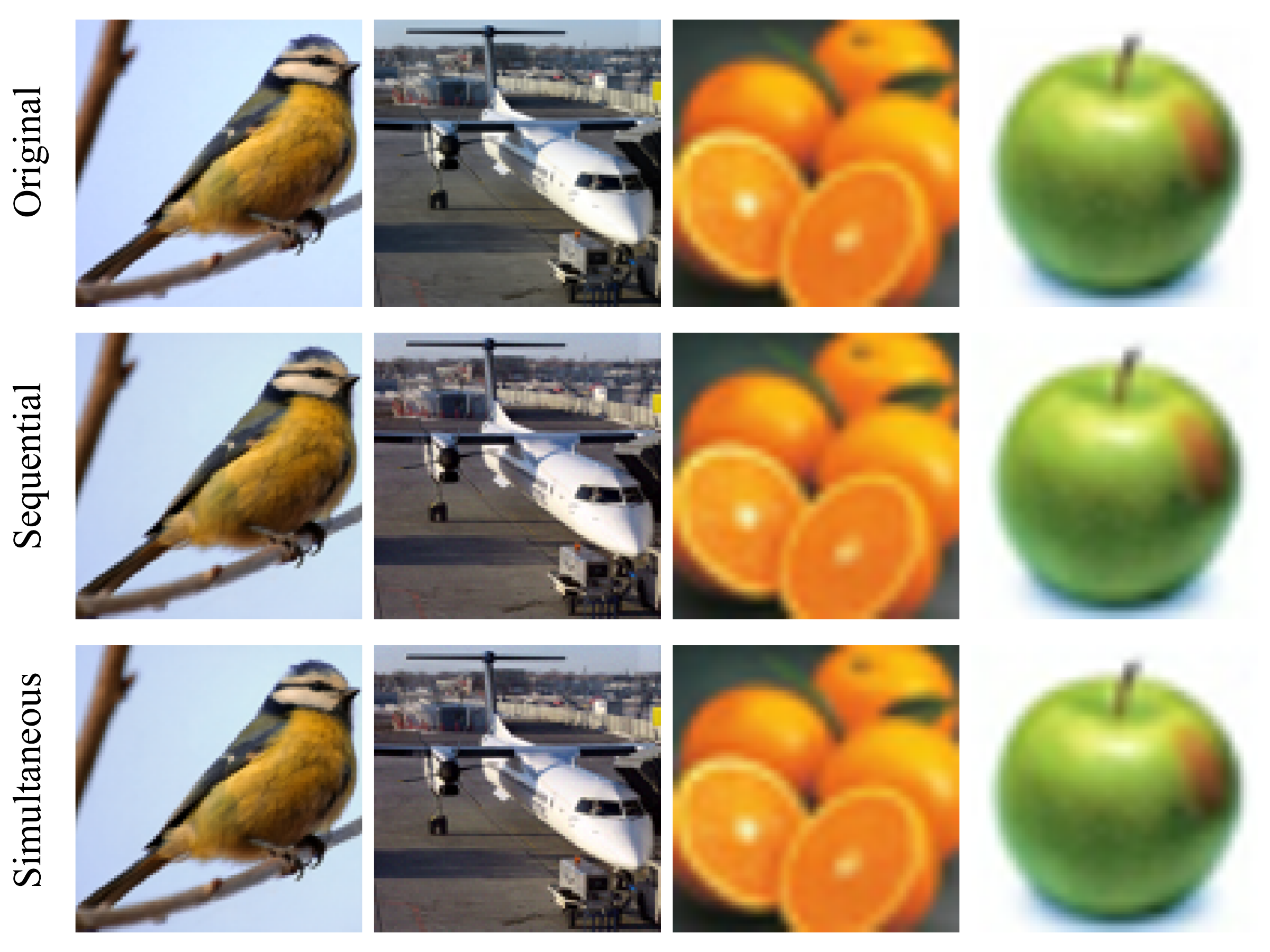}
	\caption{Demonstration of results for two multimodal-supervised semantic expressions and two selection mechanisms.}
	\vspace{-6pt}
	\label{fig:pltresult}
	\vspace{-4pt}
\end{figure}

In order to evaluate the universality of the system, we randomly select an image from the 'airplane' category of the STL-10 dataset for transmission experiments. Additionally, we choose one image each from the 'orange' and 'apple' categories of the CIFAR-100 dataset to assess the impact of low-resolution images. The SNR is set to be 10 dB in the AWGN channel. We conduct simulations employing two multimodal semantic transfer methods, along with two corresponding selection mechanisms. The results are illustrated in Fig. \ref{fig:pltresult}. The label "original" on the left indicates the image intended for transmission, while the images labeled "Sequential" and "Simultaneous" in the middle and right positions represent the reconstructed images corresponding to the two transmission and selection mechanisms. Visual inspection reveals that both multimodal-supervised semantic expressions yield highly satisfactory results.

\begin{figure}[tb]
	\vspace{-2pt}
	\centering
	\includegraphics[scale=0.23]{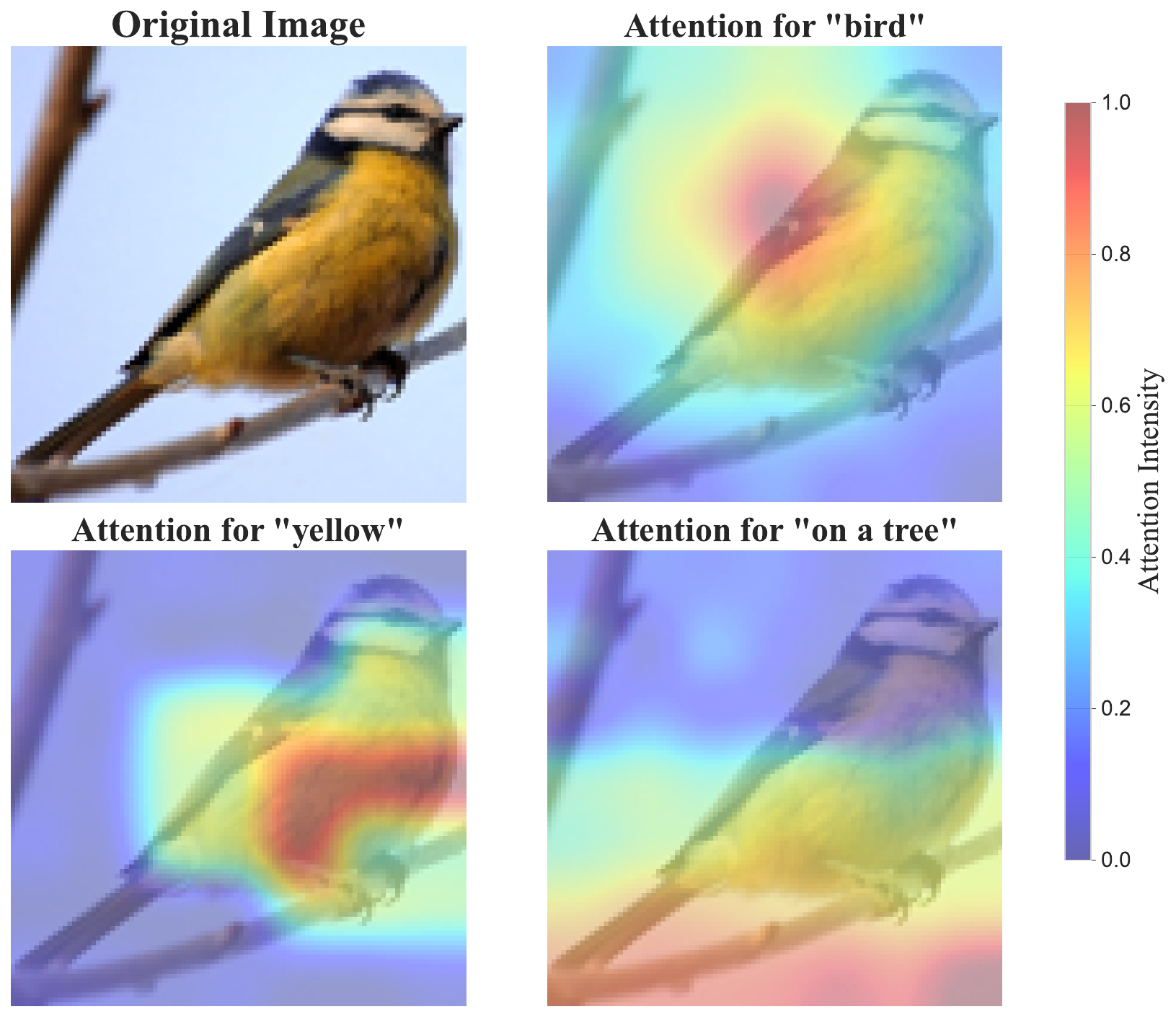}
	\caption{Weighted heat maps of text features with the proposed attention mechanism in feature fusion process.}
	\vspace{-4pt}
	\label{fig:attention_heatmap}
\end{figure}

Fig. \ref{fig:attention_heatmap} illustrates the attention heatmap of text features in relation to images within the proposed attention mechanism, specifically when text features and image features are integrated. We designate the image S in Fig. \ref{fig:system} as the target image, and select three text features as 'bird', 'yellow' and 'on a tree'. In the fusion operation with image features, the attention weights assigned to these three text features by our attention mechanism are 0.263, 0.231, and 0.223, respectively. The heatmap also indicates that the text features focus effectively on the corresponding areas of the image, ensuring that the resulting fused features accurately represent the integration of both text and image features.

\vspace{-6pt}
\subsection{Performance Comparison}
\vspace{-2pt}
In our image transmission system, we conduct a comparative analysis of two proposed multimodal semantic transmission methods against four established image codec approaches: Better Portable Graphics (BPG), Google's WebP, the Deep Joint Source-Channel Coding (DeepJSCC) system \cite{DeepJSCC}, the large language model-driven semantic communication system (LaMoSC) \cite{LaMoSC}, Wireless image transmission transformer semantic communication system (WITT) \cite{WITT} and the ablation experiments without CLIP in the MMSemCom system. We consider two image quality assessment metrics, namely PSNR and MS-SSIM. The evaluation is performed under a range of SNR conditions, spanning from 0 dB to 20 dB, while considering the effects of AWGN and Rayleigh fading. For the path loss, we consider simulations in AWGN and Rayleigh channels for transmission distances ranging from 50 m to 200 m with a transmit power of 20 dBm. PSNR serves as a quantitative measure of the reconstruction quality by calculating the mean squared error between the reconstructed and original images, with higher PSNR values indicating a superior reconstruction quality. Conversely, MS-SSIM assesses the image quality by taking into account the perceptual characteristics of the human visual system, analyzing factors such as luminance, contrast, and structural information. MS-SSIM values range from 0 to 1, with values approaching 1 indicating the improved reconstruction quality. The datasets employed in our performance comparison include CIFAR-100 and STL-10. Subsequently, we outline the four baseline methods utilized for comparison.

\begin{itemize}
	
	\item The sequential MMSemCom system involve sampling 5,000 images using a DM trained for 50,000 iterations. Then, text features and image features are employed sequentially for filtering. To assess the system's performance across datasets, we randomly select five images from the CIFAR-100 and STL-10 datasets and evaluate the average PSNR and MS-SSIM of these images under varying SNRs. This evaluation serves as a representative measure of the system's performance on these two datasets.
	
	\item The simultaneous MMSemCom system involves the sampling of 5,000 images utilizing a DM that is trained for 50,000 iterations. Subsequently, a fused multimodal feature is employed for filtering in a simultaneous manner. Additionally, we randomly select five images from the CIFAR-100 and STL-10 datasets to evaluate the average PSNR and MS-SSIM of these images under varying SNR. This evaluation serves as a representative measure of the system's performance across the two datasets.

	\item BPG, developed by Fabrice Bellard, is an image format that utilizes HEVC video compression technology. It exhibits a superior compression efficiency in comparison to JPEG, achieving smaller file sizes while preserving equivalent image quality. In this study, we employ the latest version of the BPG encoder and decoder available on the official website to encode and decode the target image at various SNRs. For the optimal performance, all parameters are set to their default values, with the exception of the quantizer hyper-parameter, which is fixed at a value of 10. Under these conditions, we calculate different values of PSNR and MS-SSIM of the target image.
	
\begin{figure*}[ht]
	\vspace{-2pt}
	\centering
	\includegraphics[scale=0.22]{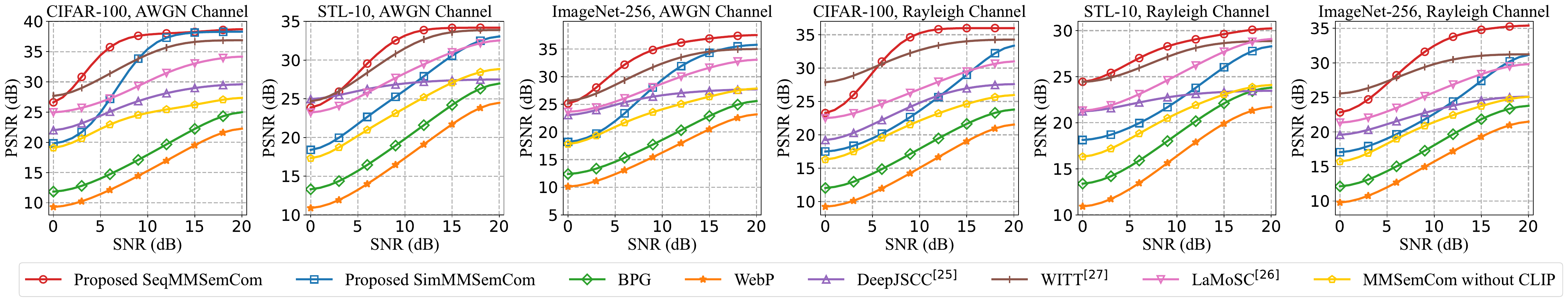}
	\caption{Comparison of the PSNR performance in different datasets with AWGN and Rayleigh fading.}
	\label{fig:psnr_comparison}
	\vspace{-4pt}
\end{figure*}

\begin{figure*}[ht]
	\centering
	\includegraphics[scale=0.22]{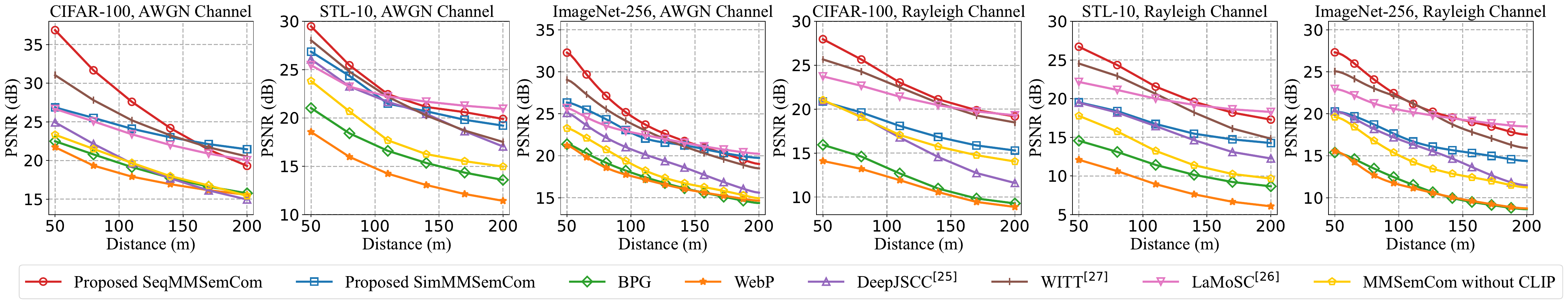}
	\caption{Comparison of the PSNR performance in different datasets for AWGN and Rayleigh fading with path loss.}
	\label{fig:psnr_pathloss_comparison}
	\vspace{-8pt}
\end{figure*}

\begin{figure*}[ht]
	\centering
	\vspace{-2pt}
	\includegraphics[scale=0.22]{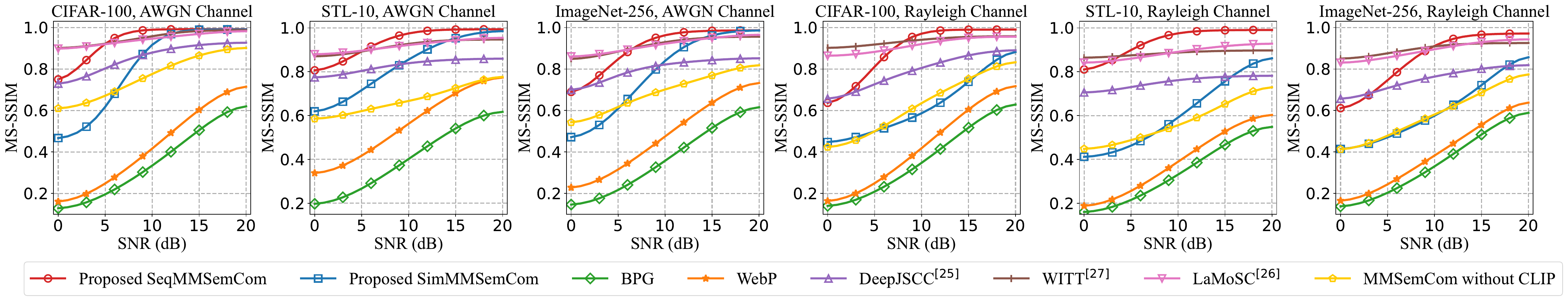}
	\caption{Comparison of the MS-SSIM performance in different datasets with AWGN and Rayleigh fading.}
	\label{fig:ssim_comparison}
	\vspace{-2pt}
\end{figure*}

\begin{figure*}[ht]
	\centering
	\includegraphics[scale=0.22]{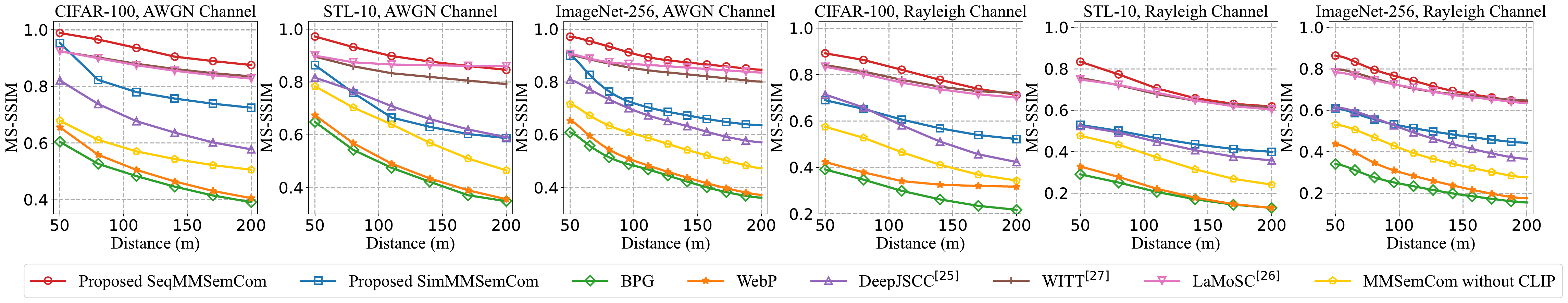}
	\caption{Comparison of the MS-SSIM performance in different datasets for AWGN and Rayleigh fading with path loss.}
	\label{fig:ssim_pathloss_comparison}
	\vspace{-4pt}
\end{figure*}

	\item WebP, developed by Google, is a contemporary image format that supports both lossy and lossless compression. This format utilizes predictive coding for image blocks and advanced arithmetic coding to achieve further compression, resulting in a significant reduction in file sizes while maintaining the image quality. We implement the WebP algorithm utilizing the WebP image compression mode available in OpenCV, with the hyperparameter quality set to 100.
	
	\item LaMoSC, presents an Large Language Model (LLM)-driven multimodal fusion semantic communication framework, which aims to expand unimodal transmission systems and enhance the generalization ability with an end-to-end encoding-decoding network that integrates visual and textual multimodal feature inputs. For our analysis, we download the checkpoint provided on GitHub \footnote[1]{https://github.com/jiangfeibo/LAMSC} and set the training epoch as 100 at a compression ratio of $1/6$ as our experimental setting. We test our two datasets on both the AWGN channel and the Rayleigh channel, recording the relevant PSNR and MS-SSIM indicators.

	\item WITT, employs Swin Transformers as an effective backbone for extracting the long-range information. It also incorporates a spatial modulation module that scales the latent representations based on the channel state information, which enhances the ability of a single model to adapt to various channel conditions. We utilize the same parameters from the code provided by the authors on GitHub \footnote[2]{https://github.com/KeYang8/WITT} to train on AWGN and Rayleigh channels using our dataset, allowing us to obtain PSNR and MS-SSIM results for comparison.

	\item DeepJSCC represents an end-to-end image transmission system that unifies source and channel coding. This approach parameterizes the encoder and decoder functions using two jointly trained CNNs, demonstrating a significant improvement over traditional separation-based digital communication methods. We conduct 1000 training iterations on the CIFAR-100 and STL-10 datasets utilizing the proposed DeepJSCC. For our analysis, we select two training settings that exhibit favorable performance as reported in the paper with a compression ratio of $1/6$ as examples. The results obtained are consistent with those presented in the original study. Additionally, we perform an evaluation of the MS-SSIM under these conditions.
\end{itemize}

\vspace{-8pt}
\subsection{Performance Analysis}

The two methods exhibit distinct differences in their sensitivity to noise. Simultaneous system relies on fusion features, which are more susceptible to noise at low SNR. This vulnerability can result in bias screening. In contrast, sequential system is conducted in stages, leading to higher redundancy, a lower probability of false selection, and greater robustness to noise due to the dual verification of text and image features. However, since the simultaneous system only needs to transmit the fused features, the communication overhead required for transmission is less than that of the sequential system that needs to transmit image features and text features.

\begin{table}[t]
	\centering
	\caption{PERFORMANCE COMPARISON OF PROPOSED METHODS UNDER DIFFERENT SNRS}
	\label{tab:snr_results}
	\begin{tabular}{@{\hspace{2pt}}c@{\hspace{2pt}}|c|c}
		\toprule
		\textbf{SNR} & \parbox{2.8cm}{\centering \textbf{SeqMMSemCom} \\ \textbf{PSNR ± Std [CI]}} & \parbox{2.8cm}{\centering \textbf{SimMMSemCom} \\ \textbf{PSNR ± Std [CI]}} \\
		\midrule
		0 dB  & 25.55 ± 0.17 [25.54, 25.56] & 19.24 ± 0.19 [19.23, 19.25] \\
		5 dB  & 32.09 ± 0.13 [32.08, 32.10] & 23.70 ± 0.13 [23.70, 23.71] \\
		10 dB & 35.62 ± 0.09 [35.61, 35.62] & 30.75 ± 0.11 [30.74, 30.75] \\
		15 dB & 36.75 ± 0.07 [36.75, 36.75] & 33.79 ± 0.06 [33.78, 33.79] \\
		20 dB & 37.29 ± 0.04 [37.29, 37.29] & 35.83 ± 0.05 [35.82, 35.83] \\
		\bottomrule
	\end{tabular}
\end{table}

To enhance the statistical validity of our conclusions, we conduct evaluations using a larger sample size and provide comprehensive statistical metrics, as shown in Table \ref{tab:snr_results}. Specifically, we randomly select 1,000 test images from each of the CIFAR-100, STL-10, and ImageNet-256 datasets to assess the reconstruction performance of our system under various SNR levels. For each configuration, we report the mean PSNR, standard deviation, and $95\%$ confidence intervals to quantify the stability and reliability of the results. This extensive sampling not only increases the representativeness of the evaluation but also enables us to verify the robustness of the proposed MMSemCom architecture across datasets with diverse semantic categories and image distributions. The observed low variance and narrow confidence intervals across different SNR levels provide strong empirical support for the claimed generalization and resilience of the system in real-world scenarios.

\subsubsection{PSNR Performance Analysis}

Fig. \ref{fig:psnr_comparison} illustrates the PSNR performance across varying SNR levels for the CIFAR-100 and STL-10 datasets in AWGN and Rayleigh fading channels. The results indicate that the proposed sequential and simultaneous MMSemCom systems consistently outperform classical coding schemes and some deep learning methods based on large language models as well as transformer architectures under different SNR conditions.

For the AWGN channel in Fig. \ref{fig:psnr_comparison}, both the sequential and simultaneous MMSemCom systems demonstrate a superior PSNR performance compared to traditional methods. In the CIFAR-100 dataset, the sequential MMSemCom system exhibits the highest performance across most of SNR levels, with a PSNR gain of approximately 4.61 dB over WITT and 7.84 dB over LaMoSC at an SNR of 6 dB. The simultaneous MMSemCom system also achieves a significant improvement, albeit slightly lower than the sequential approach. In the STL-10 dataset, the sequential MMSemCom system remains the best-performing method in the medium-to-high SNR range. Compared to LaMoSC, it achieves an improvement of 5.04 dB at an SNR of 9 dB. The simultaneous MMSemCom system shows a comparable performance to LLM based methods at high SNR values while still outperforming DeepJSCC and traditional methods, reinforcing the effectiveness of multimodal semantic feature selection in image reconstruction.

For the Rayleigh fading channel, depicted in Fig. \ref{fig:psnr_comparison}, both MMSemCom systems maintain a notable advantage over benchmarks. In the CIFAR-100 dataset, the sequential MMSemCom system achieves up to 5.31 dB improvement over BPG and 4.02 dB over DeepJSCC at high SNRs. The simultaneous MMSemCom system also demonstrates notable gains but slightly lags behind the sequential MMSemCom approach. For the STL-10 dataset in the Rayleigh channel, the sequential MMSemCom system continues to dominate in the low-to-medium SNR range. It surpasses DeepJSCC by 6.12 and LaMoSC by 3.27 dB at 9 dB and consistently outperforms all classical benchmarks in some cases. The simultaneous MMSemCom system, while slightly weaker than the sequential approach, maintains a competitive performance across all tested SNR levels.

The proposed MMSemCom system operates by filtering various semantic features at the receiver. The impact of channel noise reflects not directly on the original image signal but rather on the semantic features themselves. At the receiver end, the trained DM is capable of generating high-quality images. Consequently, if the appropriate image can be selected using the semantic features post-transmission, the PSNR value is expected to perform favorably. In the sequential MMSemCom system, the utilization of two types of semantic features for filtering enhances the robustness of the system, ensuring that satisfactory results are maintained even at the low SNR. Conversely, in the simultaneous MMSemCom system, where only the multimodal-supervised semantic fusion features are employed for screening, the adverse effects of the noise due to the low SNR become more pronounced. This results in a greater disparity between the semantic features after transmission and those prior to transmission, leading to a minor degradation in the system performance at low SNR levels. Moreover, the results of ablation experiments conducted without CLIP are significantly lower than those obtained with MMSemCom using CLIP. This effectively demonstrates the supervisory role of our multimodal features in the transmission process.

\subsubsection{MS-SSIM Performance Analysis}

Fig. \ref{fig:ssim_comparison} illustrates the MS-SSIM performance in an AWGN and Rayleigh channels across varying SNR for the CIFAR-100 and STL-10 datasets. The data presented in the figure indicates that both of our proposed MMSemCom systems exhibit significantly a superior performance compared to other benchmarks in the CIFAR-100 dataset, rapidly approaching the maximum value of 1. Notably, the sequential MMSemCom system remains close to the maximum MS-SSIM value even at low SNRs, suggesting that the images reconstructed using this method closely resemble the original graphics and consistently maintain a high performance. In contrast, the performance of the simultaneous MMSemCom system aligns with the PSNR index, exhibiting a cliff effect at low SNRs. Specifically, at an SNR of 10 dB, the simultaneous MMSemCom system demonstrates an improvement of $6.8\%$ over the Deep JSCC and a $55.9\%$ enhancement compared to WebP. The trends and performance metrics observed in the STL-10 dataset closely mirror those of the CIFAR-100 dataset, with the sequential MMSemCom system consistently outperforming other benchmarks throughout the evaluation. Conversely, the simultaneous MMSemCom system exhibits diminished performance at lower SNRs. However, the MS-SSIM value increases rapidly with the increasing SNR, approaching its maximum value.

The images produced through diffusion training closely resemble the original images. This training process involves the systematic elimination of the random noise, which enhances the quality of the sampled images generated after a specified number of training iterations. Consequently, the reconstructed images from two MMSemCom systems presented in this paper demonstrate high values on the MS-SSIM.


The significant performance gain of MMSemCom, particularly the sequential variant, over LaMoSC and WITT stems from key architectural differences, especially in handling channel noise. In typical end-to-end systems like LaMoSC, the decoder directly reconstructs the full image from noisy features. Under the low SNR, even a small deviation in semantic features can result in blurring, artifacts, or content errors in the reconstructed image, which represents a significant bottleneck in performance.

In contrast, our MMSemCom system alters the complexity of the decoding process. Leveraging its robust prior knowledge, the DM at the receiver ensures that each image in the generated candidate image set is clear and aligns with the data distribution. The impact of channel noise is primarily reflected in the selection of the image. The system's task shifts from creating a complex image based on the ambiguous semantics to selecting the best match from a set of clear images according to that same ambiguous semantics. Clearly, the latter approach is significantly more fault-tolerant. This explains why our system does not exhibit the performance cliff of other methods in low SNR conditions and can maintain a high level of robustness.

\subsubsection{Path Loss Impact Analysis}
Fig. \ref{fig:psnr_pathloss_comparison} evaluates the PSNR performance under varying transmission distances (50 m to 200 m) for both datasets in AWGN and Rayleigh channels, reflecting the effects of path loss on system performance. As the distance increases, path loss degrades the effective SNR, leading to a progressive decline in the reconstruction quality across all methods.

For CIFAR-100 dataset in AWGN, sequential MMSemCom maintains a PSNR above 25 dB even at 100 m, outperforming WITT by 3.1 dB and WebP by 8.7 dB. The sequential system’s dual semantic filtering mechanism mitigates path loss-induced distortions, retaining 57$\%$ of its maximum PSNR at 200 m. In contrast, classical codecs  exhibit rapid degradation, losing over 40$\%$ of their peak PSNR at 200 m. For STL-10 in AWGN, sequential MMSemCom achieves the 20 dB PSNR at 200 m, surpassing DeepJSCC by 4.9 dB and WITT by 2.3 dB. SimMMSemCom shows moderate robustness, maintaining a 7.4 dB advantage over WebP at 200 m.

In Rayleigh channels with path loss, the sequential system demonstrates the exceptional resilience. For CIFAR-100, the sequential MMSemCom retains the 19.6 dB PSNR at 200 m, outperforming BPG by 10 dB and DeepJSCC by 7.2 dB. Notably, the performance gap between the sequential MMSemCom and simultaneous MMSemCom widens to 4.5 dB at 200 m, emphasizing the importance of sequential feature refinement in combating combined fading and path loss. For STL-10, the sequential MMSemCom sustains 17.6 dB PSNR at 200 m, while traditional codecs fall below 10 dB due to their inability to adapt to dynamic channel conditions.

The MS-SSIM metrics in Fig. \ref{fig:ssim_pathloss_comparison} further validate these trends. At 200 m in Rayleigh channels, the sequential MMSemCom maintains MS-SSIM over 0.7 for CIFAR-100 and 0.6 for STL-10, whereas DeepJSCC drops to 0.54 and 0.4.

These results underscore that path loss exacerbates the challenges posed by channel noise and fading. The sequential MMSemCom’s hierarchical semantic feature prioritization ensures stable performance across distances, while classical methods suffer from significant quality degradation, particularly beyond 150 m. This analysis reinforces the necessity of semantic-aware adaptive systems for long-range wireless visual communications.

\subsubsection{Robustness and Generalizability}

To further examine the generalization capability of the proposed model, we evaluate it on out-of-distribution (OOD) datasets that share semantic categories with the training data. Specifically, we assess the model's generalization performance using the FGVC-Aircraft dataset, which contains fine-grained images of aircraft captured from various angles and perspectives, as well as the Stanford-Cars dataset, which includes high-resolution images of cars from different makes and models. These cross-dataset evaluations enable us to test the robustness of the semantic communication system when confronted with domain shifts and to investigate whether the learned semantic representations are transferable across datasets with varying visual statistics and complexity.

\begin{figure}[tb]
	\vspace{-2pt}
	\centering
	\includegraphics[scale=0.19]{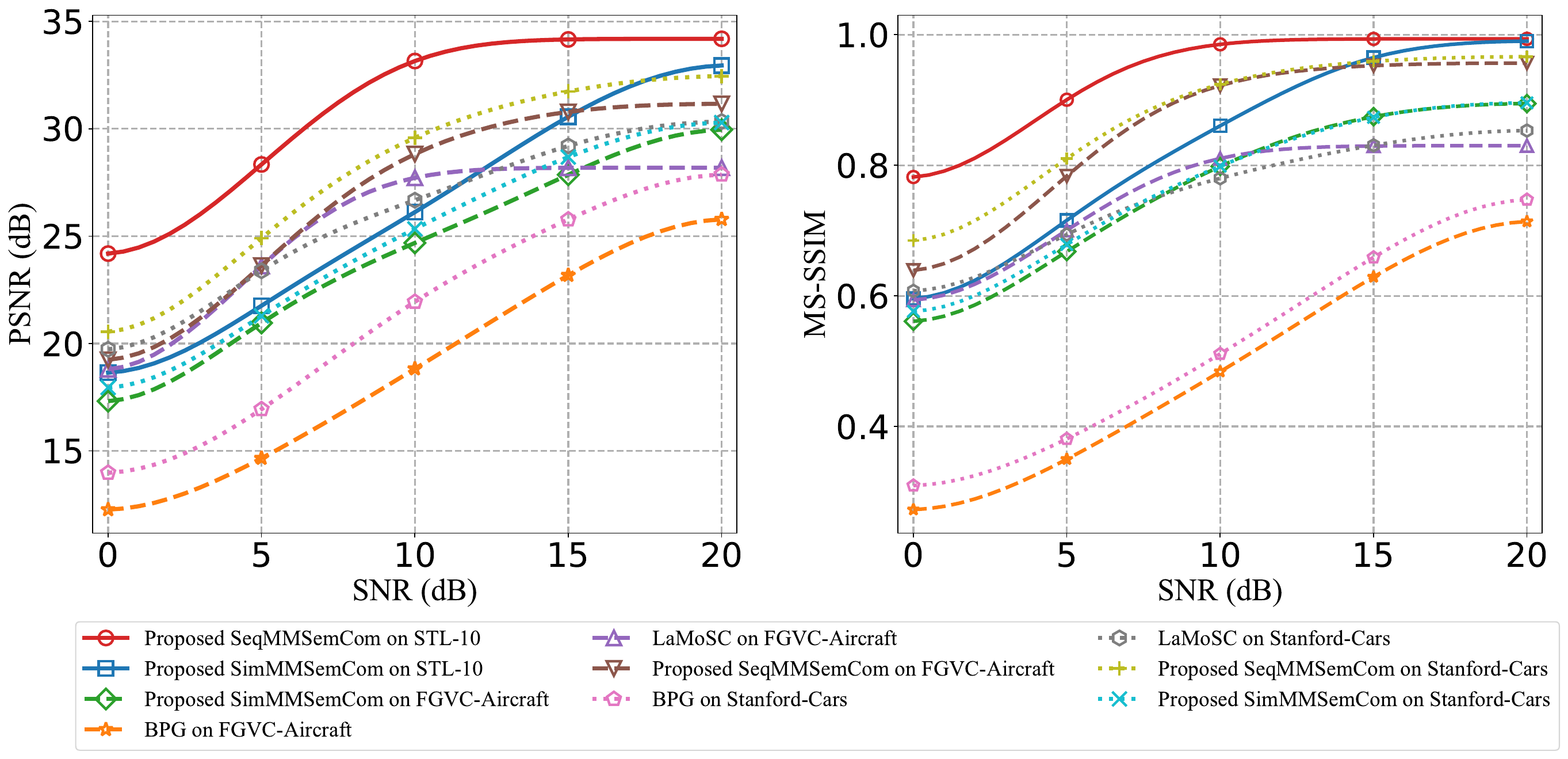}
	\caption{The generalization performance of the trained model in the out-of-distribution datasets.}
	\vspace{-6pt}
	\label{fig:gene}
	\vspace{-8pt}
\end{figure}

As illustrated in Fig. \ref{fig:gene}, our MMSemCom systems demonstrate remarkable generalization performance. When tested on the unseen FGVC-Aircraft and Stanford-Cars datasets, both SeqMMSemCom and SimMMSemCom exhibit only a minor performance degradation compared to their baseline results on the in-distribution STL-10 data. More importantly, they maintain a significant performance advantage over competing methods such as BPG and LaMoSC, even in these challenging OOD scenarios. This sustained advantage strongly indicates that our framework possesses robust generalization capabilities, which can be attributed to the universal nature of the extracted multimodal features and the effectiveness of our generation and selection paradigm in bridging the domain gap.

\begin{table}[htbp]
	\centering
	\caption{PSNR PERFORMANCE COMPARISON OF PROPOSED METHODS UNDER DIFFERENT SNRS AND PROMPT QUANTITIES}
	\label{tab:psnr_comparison}
	\begin{tabular}{@{\hspace{3pt}}c@{\hspace{3pt}}|@{\hspace{3pt}}c@{\hspace{3pt}}|c|c|c|c|cc}
			\toprule
			\multirow{2}{*}{Method} & \multirow{2}{*}{\begin{tabular}[c]{@{}c@{}}Prompt\\Quantity\end{tabular}} & \multicolumn{5}{c}{SNR (dB)} \\
			\cmidrule(lr){3-7}
			& & 0 & 5 & 10 & 15 & 20 \\
			\midrule
			\multirow{4}{*}{SeqMMSemCom} 
			& 1 & 20.93 & 23.31 & 26.04 & 29.11 & 29.99 \\
			& 3 & 24.73 & 28.12 & 33.16 & 34.29 & 34.32 \\
			& 5 & 25.47 & 29.02 & 33.48 & 34.56 & 34.61 \\
			& 7 & 25.68 & 29.34 & 33.51 & 34.71 & 34.82 \\
			\cmidrule(lr){1-7}
			\multirow{4}{*}{SimMMSemCom} 
			& 1 & 14.65 & 18.48 & 21.97 & 25.47 & 29.32 \\
			& 3 & 18.47 & 21.97 & 25.75 & 30.02 & 33.84 \\
			& 5 & 19.64 & 23.13 & 26.98 & 31.43 & 34.18 \\
			& 7 & 20.08 & 23.51 & 27.25 & 31.84 & 34.43 \\
			\bottomrule
	\end{tabular}

\end{table}

To quantify the impact of textual semantic information on the reconstruction quality, we conduct an ablation study examining the quantity of prompts as shown in Table \ref{tab:psnr_comparison}. Utilizing the trained model on the STL-10 dataset, we systematically vary the number of prompts from 1 to 7 and measure the corresponding PSNR.
	
The results presented in Table \ref{tab:psnr_comparison} reveal a clear trend: performance initially improves significantly as the number of prompts increases. This improvement occurs because additional prompts provide a richer and more descriptive semantic context, enabling the selection mechanism at the receiver to more accurately distinguish the target image from other candidates in the generated set. However, as the prompt quantity continues to grow, the performance gains diminish and eventually reach a saturation point. This saturation indicates that a limited number of core prompts are sufficient to capture the most salient semantics, while additional prompts may introduce redundancy or overly specific details that do not substantially enhance selection accuracy.

Crucially, each additional prompt incurs a higher cost. It increases the communication overhead, as indicated by the term $n_{text}$ in our overhead analysis, and also adds to the computational load at the receiver during the feature comparison stage. Therefore, there exists a critical trade-off between reconstruction accuracy and system efficiency. Our findings indicate that using three prompts strikes an effective balance, achieving near-optimal performance without incurring excessive communication and computational costs. This justifies our choice of $n_{text}=3$ in the main experiments.

\subsubsection{Complexity Analysis}

We present the computational complexity metrics of various models in Table \ref{table:comp}. This includes the model parameters, floating point operations (FLOPs), inference time for a single image, and the amount of GPU memory required for inference on a single image. We conduct a comparative analysis of our system model alongside three other deep learning-based models using the CIFAR-100 dataset, utilizing a single NVIDIA A100 GPU. As illustrated by the results in Table \ref{table:comp}, our designed model is 3.65 times smaller than WITT and slightly smaller than LaMoSC. Additionally, the GPU memory usage of our model is lower than that of both WITT and LaMoSC, while still achieving the performance that surpasses these two models. Although our model has more parameters and FLOPs than DeepJSCC, MMSemCom significantly outperforms DeepJSCC. Furthermore, our model boasts the fastest inference time for a single image, which clearly demonstrates its high efficiency.

\begin{table}[t]
	\centering
	\captionsetup{justification=centering} 
	\renewcommand{\arraystretch}{1.5}
	\caption{\centering COMPARISON OF COMPUTATIONAL RESOURCES OF DIFFERENT MODELS}
	\centering
	\scriptsize
	\begin{tabular}{|@{\hspace{3pt}}c@{\hspace{3pt}}|@{\hspace{3pt}}c@{\hspace{3pt}}|@{\hspace{3pt}}c@{\hspace{3pt}}|@{\hspace{3pt}}c@{\hspace{3pt}}|@{\hspace{3pt}}c@{\hspace{3pt}}|}
		\hline
		\textbf{Method} & \textbf{Parameters} & \textbf{FLOPs} & \textbf{Inference Time} & \textbf{GPU Memory}\\
		\hline
		MMSemCom & 7.71M & 96G & 35ms & 314.77MB\\
		\hline
		WITT  & 28.2M  & 198G & 116ms & 1.12GB\\
		\hline
		LaMoSC & 11.17M & 107G & 255ms & 486.48MB \\
		\hline
		DeePJSCC & 0.19M  & 9.91M & 238ms & 8.89MB\\
		\hline
	\end{tabular}
	\label{table:comp}
	\vspace{-1mm}
\end{table}

\begin{table}[t]
	\centering
	\captionsetup{justification=centering} 
	\renewcommand{\arraystretch}{1.2} 
	\caption{DEMONSTRATION OF PERFORMANCE AND EFFICIENCY TRADE-OFF OF DIFFERENT ABLATION METHODS}
	\label{tab:tradeoff_final}
	\scriptsize 
	\begin{tabular}{|@{\hspace{3pt}}c@{\hspace{3pt}}|@{\hspace{3pt}}c@{\hspace{3pt}}|@{\hspace{3pt}}c@{\hspace{3pt}}|@{\hspace{3pt}}c@{\hspace{3pt}}|@{\hspace{3pt}}c@{\hspace{3pt}}|}
		\hline
		\rule{0pt}{4ex}\textbf{Mode} & \makecell{\textbf{Overhead} \\ \textbf{(Bytes)}} & \makecell{\textbf{PSNR} \\ \textbf{(dB)}} & \textbf{MS-SSIM} & \makecell{\textbf{Efficiency} \\ \textbf{(PSNR/KB)}} \\
		\hline
		\rule{0pt}{2.5ex}Image Only   & 128    & 14.526 & 0.414 & 116.208 \\
		\hline
		\rule{0pt}{2.5ex}Text Only    & 6144   & 21.872 & 0.535 & 3.645   \\
		\hline
		\rule{0pt}{2.5ex}Simultaneous & 256    & 30.745 & 0.853 & 122.98  \\
		\hline
		\rule{0pt}{2.5ex}Sequential   & 6272   & 35.616 & 0.996 & 5.815   \\
		\hline
	\end{tabular}
	\vspace{-1mm} 
\end{table}

A quantitative analysis of the trade-off between the reconstruction quality and the communication overhead is presented in Table \ref{tab:tradeoff_final}, with the performance evaluated under fixed channel conditions. The results clearly illustrate the limitations of single-modality methods: the "Image Only" mode yields a low PSNR of 14.526 dB, while the "Text Only" mode is highly inefficient, requiring over 6 KB for only a marginal improvement in the quality. In contrast, our "Sequential" approach establishes the performance ceiling at 35.616 dB PSNR, albeit at the highest communication cost. Most notably, the "Simultaneous" mode achieves a high-quality reconstruction of 30.745 dB PSNR with an overhead of only 256 bytes. This exceptional balance is quantitatively captured by its top-ranking efficiency of 122.98 PSNR/KB, validating our attention-based fusion as a practical and effective approach that optimally balances high fidelity with minimal communication costs.

\vspace{-8pt}
\subsection{Demonstration of a multi-user scenario}
\vspace{-8pt}

\begin{figure}[ht]
	\vspace{-2pt}
	\centering
	\includegraphics[scale=0.24]{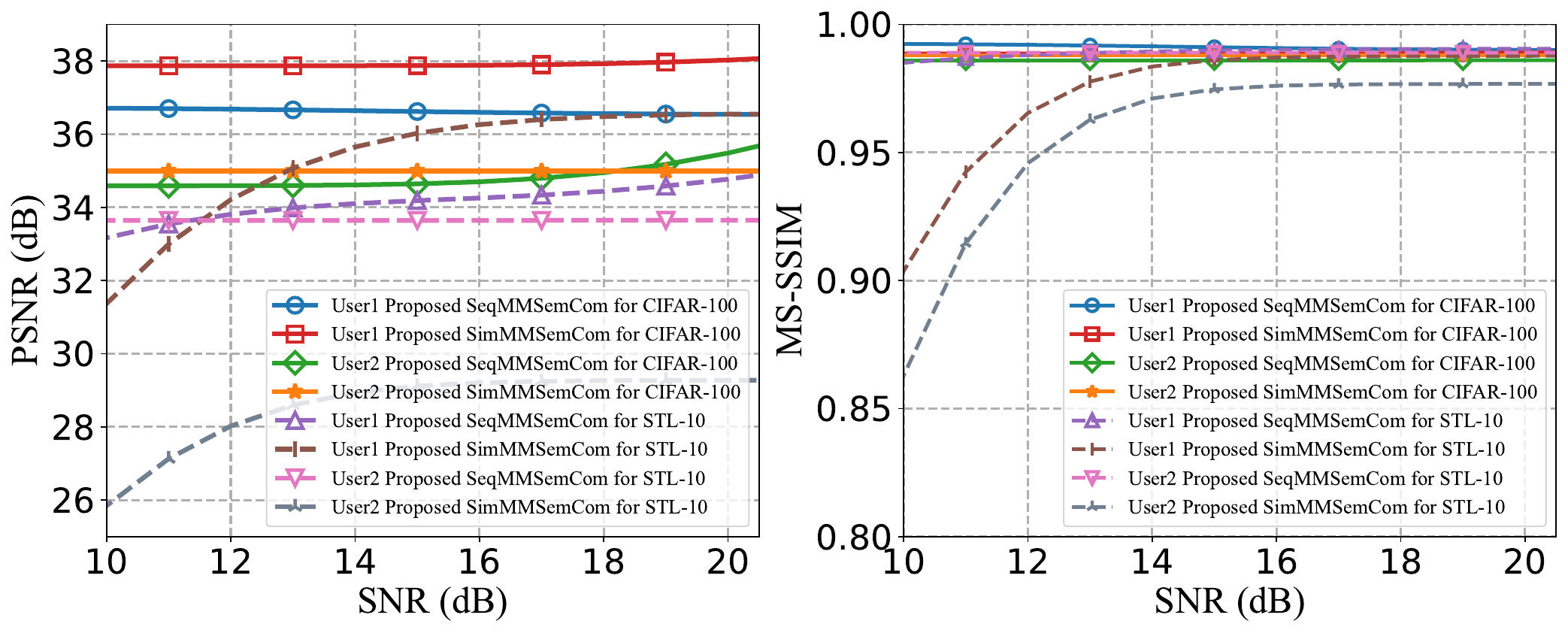}
	\caption{PSNR and MS-SSIM performance of multiuser scenarios on CIFAR-100 and STL-10 datasets for two MMSemCom systems.}
	\label{fig:psnrmultiuser}
	\vspace{-2pt}
\end{figure}

We also evaluate the performance of our two MMSemCom systems in a multi-user scenario, as outlined in our Walsh code transmission framework proposed in Section \ref{sec:Multi-User Scenario}. For this evaluation, we select two images, one from the CIFAR-100 dataset and another from the STL-10 dataset, to be transmitted to two users under AWGN channel. Besides, we employ the Hadamard transform. The matrix generates a block Walsh code matrix that corresponds to the dimensions of the text prompts and image features, encoding the necessary semantic features for transmission as described in eqs. (22) and (23) at the transmitter. Additionally, it facilitates the decoding of text and image features at the receiver, as outlined in  eqs. (24) through  eqs. (26). We evaluate the PSNR and MS-SSIM performance for User 1 and User 2 across various SNRs, as illustrated in Fig. \ref{fig:psnrmultiuser}. The figures indicate that when the SNR exceeds 10 dB, both users can achieve an enhanced performance for both datasets. Conversely, at SNR levels below 12 dB, the performance evaluated using the STL-10 dataset, which contains high pixel values, deteriorates significantly, demonstrating a pronounced cliff effect. This decline in the performance is attributed to the substantial impact of low SNR on the semantic features, which hampers the effective provision of selection references post-transmission, ultimately leading to suboptimal selection outcomes.

\vspace{-2pt}
\section{Conclusion} 
\label{sec:conclusion}

In this paper, we propose a framework for multimodal-supervised generative image semantic communication and introduce two MMSemCom systems, sequential MMSemCom and simultaneous MMSemCom. We delineate two methodologies for expressing multimodal semantics, accompanied by two different selection mechanisms. Additionally, we propose a multiuser transmission mode for this framework. Simulation results demonstrate that the proposed systems outperform various baseline models across different SNRs and diverse datasets. Notably, the sequential MMSemCom exhibits exceptional transmission performance at low SNRs, indicating remarkable robustness. These findings substantiate the potential of our proposed generation and selection framework and its innovative approach to semantic information representation, highlighting its applicability in image semantic communication systems and a broad spectrum of application scenarios.

Future work will focus on addressing these limitations. A promising direction involves transitioning from class-conditional generation to a more universal text-to-image synthesis paradigm, which will enhance the flexibility and scalability. We will also investigate more efficient generative models and early-exit selection strategies to reduce the computational load on the receiver. Finally, exploring mechanisms for the dynamic updating of the shared knowledge base will be crucial for adapting the system to evolving real-world environments.

\ifCLASSOPTIONcaptionsoff
\newpage
\fi

\vspace{-4pt}
\bibliographystyle{IEEEtran}
\vspace{-2pt}
\bibliography{refer.bib}

\end{document}